\documentclass{aa}  
\usepackage{graphicx}
\usepackage{color}
\usepackage{natbib}
\usepackage{textcomp}
\usepackage{tabulary}
\usepackage{multirow}
\usepackage{siunitx}
\usepackage{txfonts}
\usepackage[T1]{fontenc}
\newcounter{RomC}

\begin{document}

   \title{The Solar Orbiter SPICE instrument}
   \subtitle{An extreme UV imaging spectrometer}
   
   \author{SPICE Consortium: 
M.~Anderson      \inst{\ref{inst:ral}},
T.~Appourchaux   \inst{\ref{inst:ias}},
F.~Auch\`ere     \inst{\ref{inst:ias}},
R.~Aznar Cuadrado\inst{\ref{inst:mps}},
J.~Barbay        \inst{\ref{inst:ias}},
F.~Baudin        \inst{\ref{inst:ias}},
S.~Beardsley     \inst{\ref{inst:ral}},
K.~Bocchialini   \inst{\ref{inst:ias}},
B.~Borgo         \inst{\ref{inst:ias}},
D.~Bruzzi        \inst{\ref{inst:ral}},
E.~Buchlin       \inst{\ref{inst:ias}},
G.~Burton        \inst{\ref{inst:ral}},
V.~B\"uchel	     \inst{\ref{inst:pmod}},
M.~Caldwell	     \inst{\ref{inst:ral}},
S.~Caminade      \inst{\ref{inst:ias}},
M.~Carlsson      \inst{\ref{inst:uio}},
W.~Curdt         \inst{\ref{inst:mps}},
J.~Davenne       \inst{\ref{inst:ral}},
J.~Davila        \inst{\ref{inst:gsfc}},
C.~E.~DeForest   \inst{\ref{inst:swri}},
G.~Del Zanna    \inst{\ref{inst:cam}},
D.~Drummond      \inst{\ref{inst:ral}},
J.~Dubau         \inst{\ref{inst:ias}},
C.~Dumesnil      \inst{\ref{inst:ias}},
G.~Dunn          \inst{\ref{inst:swrisa}},
P.~Eccleston     \inst{\ref{inst:ral}},
A.~Fludra        \inst{\ref{inst:ral}},
T.~Fredvik       \inst{\ref{inst:uio}},
A.~Gabriel       \inst{\ref{inst:ias}},
A.~Giunta        \inst{\ref{inst:ral}},
A.~Gottwald      \inst{\ref{inst:ptb}},
D.~Griffin       \inst{\ref{inst:ral}},
T.~Grundy        \inst{\ref{inst:ral}},
S.~Guest         \inst{\ref{inst:ral}},
M.~Gyo           \inst{\ref{inst:pmod}},
M.~Haberreiter   \inst{\ref{inst:pmod}},
V.~Hansteen      \inst{\ref{inst:uio}},
R.~Harrison      \inst{\ref{inst:ral}},
D.~M.~Hassler    \inst{\ref{inst:swri}},
S.~V.~H.~Haugan  \inst{\ref{inst:uio}},
C.~Howe          \inst{\ref{inst:ral}},
M.~Janvier       \inst{\ref{inst:ias}},
R.~Klein         \inst{\ref{inst:ptb}},
S.~Koller        \inst{\ref{inst:pmod}},
T.~A.~Kucera 	\inst{\ref{inst:gsfc}},
D.~Kouliche      \inst{\ref{inst:ias},\ref{inst:cesamseed}},
E.~Marsch        \inst{\ref{inst:cau}},
A.~Marshall      \inst{\ref{inst:ral}},
G.~Marshall      \inst{\ref{inst:ral}},
S.~A.~Matthews      \inst{\ref{inst:ucl}},
C.~McQuirk       \inst{\ref{inst:ral}},
S.~Meining       \inst{\ref{inst:mps}},
C.~Mercier       \inst{\ref{inst:ias}},
N.~Morris        \inst{\ref{inst:ral}},
T.~Morse         \inst{\ref{inst:ral}},
G.~Munro         \inst{\ref{inst:esr}},
S.~Parenti       \inst{\ref{inst:ias}},
C.~Pastor-Santos \inst{\ref{inst:ral}},
H.~Peter         \inst{\ref{inst:mps}},
D.~Pfiffner      \inst{\ref{inst:pmod}},
P.~Phelan        \inst{\ref{inst:swrisa}},
A.~Philippon     \inst{\ref{inst:ias}},
A.~Richards      \inst{\ref{inst:ral}},
K.~Rogers        \inst{\ref{inst:ral}},
C.~Sawyer        \inst{\ref{inst:ral}},
P.~Schlatter     \inst{\ref{inst:pmod}},
W.~Schmutz       \inst{\ref{inst:pmod}},
U.~Sch\"uhle     \inst{\ref{inst:mps}},
B.~Shaughnessy   \inst{\ref{inst:ral}},
S.~Sidher        \inst{\ref{inst:ral}},
S.~K.~Solanki    \inst{\ref{inst:mps}, \ref{inst:KHU}},
R.~Speight       \inst{\ref{inst:ral}},
M.~Spescha       \inst{\ref{inst:pmod}},
N.~Szwec         \inst{\ref{inst:ias}},
C.~Tamiatto      \inst{\ref{inst:ias}},
L.~Teriaca       \inst{\ref{inst:mps}},
W.~Thompson      \inst{\ref{inst:adnet}}, 
I.~Tosh          \inst{\ref{inst:ral}},
S.~Tustain       \inst{\ref{inst:ral}},
J.-C.~Vial       \inst{\ref{inst:ias}},
B.~Walls         \inst{\ref{inst:swrisa}},
N.~Waltham       \inst{\ref{inst:ral}},
R.~Wimmer-Schweingruber\inst{\ref{inst:cau}},
S.~Woodward      \inst{\ref{inst:ral}},
P.~Young		\inst{\ref{inst:gsfc}, \ref{inst:nuu}},
A.~De~Groof      \inst{\ref{inst:esac}},
A.~Pacros        \inst{\ref{inst:estec}},
D.~Williams      \inst{\ref{inst:esac}},
D.~M\"uller      \inst{\ref{inst:estec}}\thanks{Corresponding author: D. M\"uller, e-mail: Daniel.Mueller@esa.int}
   }

\authorrunning{SPICE Consortium}

   \institute{
RAL Space, STFC Rutherford Appleton Laboratory, Harwell, Didcot, OX11 0QX, UK\label{inst:ral}
\and
Institut d'Astrophysique Spatiale, 91405 Orsay Cedex, France\label{inst:ias}
\and
Max-Planck-Institut f\"ur Sonnensystemforschung, Justus-von-Liebig-Weg 3, 37077 G\"ottingen, Germany\label{inst:mps}
\and
PMOD/WRC, Dorfstrasse 33, 7260 Davos Dorf, Switzerland\label{inst:pmod}
\and
Institute of Theoretical Astrophysics, University of Oslo, P.O. Box 1029 Blindern, 0315 Oslo, Norway\label{inst:uio}
\and
NASA Goddard Space Flight Center, Greenbelt, MD, USA\label{inst:gsfc}
\and
Southwest Research Institute, 1050 Walnut Street, Boulder, CO, USA\label{inst:swri}
\and
DAMTP, Centre for Mathematical Sciences, University of Cambridge Wilberforce Road Cambridge CB3 0WA, UK\label{inst:cam}
\and
Southwest Research Institute, 6220 Culebra Rd, San Antonio, TX, USA\label{inst:swrisa}
\and
Physikalisch-Technische Bundesanstalt, Abbestra{\ss}e 2--12, 10587 Berlin, Germany\label{inst:ptb}
\and
University College London, Mullard Space Science Laboratory, Holmbury St. Mary, Dorking, Surrey, RH5 6NT, UK\label{inst:ucl}
\and
ESR Technology Ltd, 202 Cavendish Place, Birchwood Park, Warrington, Cheshire, WA3 6WU, UK\label{inst:esr}
\and
Division for Extraterrestrial Physics, Institute for Experimental and Applied Physics (IEAP), Christian Albrechts University at Kiel, Leibnizstr. 11, 24118 Kiel, Germany\label{inst:cau}
\and
European Space Agency, ESAC, Camino Bajo del Castillo s/n, Urb. Villafranca del Castillo, 28692 Villanueva de la Ca\~nada, Madrid, Spain\label{inst:esac} 
\and
European Space Agency, ESTEC, P.O. Box 299, 2200 AG Noordwijk, The Netherlands\label{inst:estec}                            
\and
ADNET Systems, Inc., Lanham, MD, USA\label{inst:adnet}
\and
CESAM SEED, 52B Bd Saint-Jacques, 75014 Paris\label{inst:cesamseed}
\and
School of Space Research, Kyung Hee University, Yongin, Gyeonggi-Do, 446-701, Republic of Korea\label{inst:KHU}
\and
Northumbria University, Newcastle Upon Tyne, NE1 8ST, UK\label{inst:nuu}
}

   \date{Received 29 March 2019 / Accepted 19 August 2019}

 
  \abstract
{}
   {The Spectral Imaging of the Coronal Environment (SPICE) instrument is a high-resolution imaging spectrometer operating at extreme ultraviolet (EUV) wavelengths. In this paper, we present the concept, design, and pre-launch performance of this facility instrument on the ESA/NASA Solar Orbiter mission.}
%
   {The goal of this paper is to give prospective users a better understanding of the possible types of observations, the data acquisition, and the sources that contribute to the instrument's signal.
}
   {The paper discusses the science objectives, with a focus on the SPICE-specific aspects, before presenting the instrument's design, including optical, mechanical, thermal, and electronics aspects. This is followed by a characterisation and calibration of the instrument's performance. The paper concludes with descriptions of the operations concept and data processing.}
   {The performance measurements of the various instrument parameters meet the requirements derived from the mission's science objectives. The SPICE instrument is ready to perform measurements that will provide vital contributions to the scientific success of the Solar Orbiter mission.}

   \keywords{Sun: UV radiation -- Sun: transition region -- Sun: corona -- Instrumentation: spectrographs -- Techniques: imaging spectroscopy -- Methods: observational}

   \maketitle

\section{Introduction}
\label{sect-intro}
The Solar Orbiter mission \citep{Mueller:2013a, Mueller2019a}, scheduled to launch in February 2020, will study the Sun and inner heliosphere with a set of remote-sensing instruments observing the Sun and solar corona and a set of in-situ instruments measuring the solar wind around the spacecraft. Together, the ten Solar Orbiter instruments will provide a complete description of the plasma making up the solar wind -- its origin, transport and composition -- vastly improving on the Helios mission \citep{Schwenn:1990aa} launched in 1974. Solar Orbiter reaches a minimum perihelion of 0.28\,AU after a series of gravity assists from Venus and Earth, which will also raise the inclination of the orbital plane to above $30^\circ$ from the ecliptic plane \citep{Garcia2019}. 
The Solar Orbiter minimum perihelion of 0.28\,AU is very similar to the Helios perihelion of 0.3\,AU, but combined with its unique out-of-ecliptic vantage point, Solar Orbiter will be able to address a fundamental question of solar physics: How does the Sun create and control the heliosphere?

Solar Orbiter will combine in-situ measurements with high-resolution remote-sensing observations of the Sun in a systemic approach to resolve fundamental science problems needed to achieve this objective. These problems include the sources of the solar wind, the causes of eruptive releases of plasma and magnetic field from the Sun known as coronal mass ejections (CMEs), the evolution of CMEs and their interaction with the ambient solar wind flow, and the origins, acceleration mechanisms and transport of solar energetic particles that may be hazardous to both human explorers and robotic spacecraft that operate in the highly variable environment outside of Earth's magnetosphere.

While essential to meeting Solar Orbiter's scientific objectives, the mission's orbit also poses specific challenges to the remote-sensing instruments. For example, the changing distances to Sun and Earth result in large variations of the thermal conditions and telemetry rates along each orbit, respectively. The strategies devised jointly by the remote-sensing instruments to alleviate these constraints are described in \cite{Auchere2019a}.

\begin{figure}
\includegraphics[width=\columnwidth]{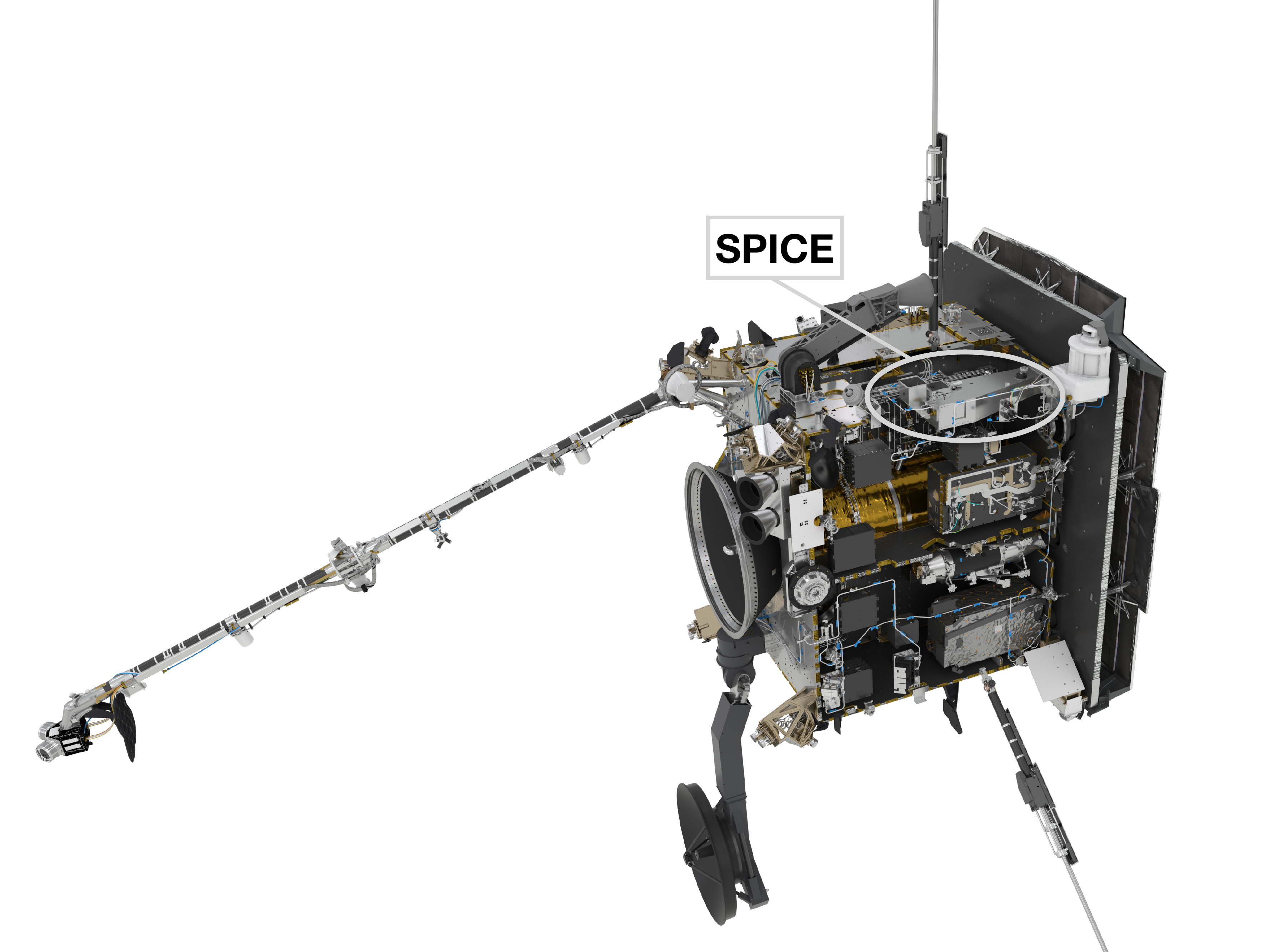}

\caption{Solar Orbiter spacecraft, with parts of the side panels removed to show the SPICE instrument.}
\label{fig:so_sc}
\end{figure}

The SPICE instrument (Fig.~\ref{fig:so_sc}) is a high-resolution imaging spectrometer operating at extreme ultraviolet (EUV) wavelengths from 70.4\,nm$-$79.0\,nm and 97.3\,nm$-$104.9\,nm. It is a facility instrument on the Solar Orbiter mission, funded by ESA member states and ESA. SPICE is allocated 45.3\,Gbits of data per six-month orbit, to be acquired nominally during three ten-day remote-sensing windows, which corresponds to an average data rate of 17.5\,kbit\,s$^{-1}$. 
Most scientific objectives do not require downloading of full spectra but only selected windows centred on typically ten spectral lines of interest. Further reduction of the data volume can be obtained either by data compression or by computing on board the total intensity of the lines. The allocated resources do not impose compressing the data beyond a ratio of 20:1 (Sect.~\ref{sec:compression}). In Sect.~\ref{sect-ops}, we provide examples of observations that illustrate the ability of SPICE to operate within the Solar Orbiter constraints.

SPICE will address the key science goals of Solar Orbiter by providing the quantitative knowledge of the physical state and composition of the plasma in the solar atmosphere, in particular investigating the source regions of outflows and ejection processes that link the solar surface and corona to the heliosphere.
SPICE is of particular importance for establishing the link between remote-sensing and in-situ measurements as it is uniquely capable of remotely characterising the plasma properties of source regions, which can directly be compared with in-situ measurements taken by the Solar Wind Analyser (SWA) instrument suite \citep{Owen2019a}.  
In magnetically closed regions, SPICE will play an essential role in characterising the turbulent state of the plasma over a wide range of temperatures from the chromosphere into the hottest parts of the corona. This is essential to understand which processes heat the plasma and drive the dynamics we observe, be it through waves, field-line braiding, or reconnection.
                                                                          
\section{Scientific objectives and opportunities}
\label{sect-obj}
The main science goals of SPICE are related to our understanding of the complex dynamic connection between the Sun and the inner heliosphere.
In this sense, the scientific focus is on studies that combine the remote-sensing and the in-situ instruments on Solar Orbiter to work as one comprehensive suite.
At the same time, the unique instrumental capabilities of SPICE will also allow stand-alone studies that will address other unsolved problems in solar physics.

By observing the intensities of selected spectral lines and their spectral profiles, SPICE will allow the temperature, density, flow, elemental composition and the turbulent state of the plasma in the upper solar atmosphere to be characterised. Emission lines originating between the top of the chromosphere and the low corona cover the temperature range from 10,000\,K to 2\,MK, augmented by two 10\,MK lines seen in flaring plasma (see Table\,\ref{T.lines}).

Following a discussion of the observables provided by SPICE in Sect.~\ref{S.observable}, we present a selection of scientific topics that will be addressed by SPICE (Sect.~\ref{S.all.instruments}).  
Naturally, this list will be incomplete, but should give a flavour of the scientific opportunities provided by SPICE.

\begin{table}
\caption{Selection of spectral lines covered by SPICE.\label{T.lines}}
\begin{tabular}{l@{}l@{~~}r@{}l@{~~}crrr}
\hline\hline
&&&&&\\
       & Ion              & $\lambda$ [\AA] && $\log{T}$\,[K] & \multicolumn{3}{c}{Intensity [ph.pix$^{-1}$.s$^{-1}$]}\\
       &                  &                 &&                & AR & QS& Ref.\\
\hline
$\star$ & \ion{H}{I}      & 1025.72 &       & 4.0 & 883.5 & 372.2 & [1]\\
        & \ion{C}{II}     & 1036.34 &       & 4.3 & 17.6 & 21.1 & [1]\\
$\star$ & \ion{C}{III}    & 977.03  &       & 4.5 & 563.7 & 312.1 & [1]\\
        & \ion{S}{V}      & 786.47  &       & 5.2 & 31.6 & 3.4 & [1]\\
        & \ion{O}{IV}     & 787.72  &       & 5.2 & 56.1 & 6.0 & [1]\\
        & \ion{O}{V}      & 760.43  &       & 5.4 & 59.5 & 1.9 & [1]\\
$\star$ & \ion{O}{VI}     & 1031.93 &       & 5.5 & 8268.2 & 139.0 & [1]\\
$\star$ & \ion{O}{VI}     & 1037.64 &       & 5.5 & 2951.3 & 66.19& [1]\\
        & \ion{Ne}{VI}    & 1005.79 &       & 5.6 & 15.4 & 0.3& [1]\\
        & \ion{Si}{VII}   & 1049.25 &       & 5.6 & 7.5 & -& [1]\\
$\star$ & \ion{Ne}{VIII}  & 770.42  &       & 5.8 & 63.9 & 7.8& [1]\\
        & \ion{Mg}{VIII}  & 772.31  &       & 5.9 & 9.2 & -& [1]\\
        & \ion{Mg}{IX}    & 706.02  &       & 6.0 & 3.0 & 0.9& [1]\\
        & \ion{Fe}{X}     & 1028.04 &       & 6.0 & 10.1 & 4.7& [1]\\
        & \ion{Mg}{XI}    & 997.44  &       & 6.2 & 1.7 & 0.6& [1]\\
$\star$ & \ion{Si}{XII}   & 520.67  &$^\dag$& 6.3 & 31.2 & 2.5 & [2]\\
        & \ion{Fe}{XVIII} & 974.84  &       & 6.9 & 6.9 & -& [2]\\
$\star$ & \ion{Fe}{XX}    & 721.55  &       & 7.0 & 1428.2 & -& [2]\\
\hline
\end{tabular}
\tablefoot{For each line the rest wavelength, $\lambda$, the approximate logarithmic line formation temperature, $T$, and the
number of photons detected by SPICE for active regions (AR) and quiet Sun (QS), are listed. The last column provides the reference for the solar flux used to simulate SPICE observations. 
The intensities are for the 2\arcsec slit binned over two spatial pixels.
Lines marked with a $\star$ are strong lines for which full line profiles can routinely be returned.
For the weaker lines, the intensity integrated across the line will be computed on-board and sent down. 
The \ion{Fe}{XX} intensity is from a M7.6 flare. 
\ion{Si}{XII}, marked by $^\dag$, will be observed in the second spectral order. This line and \ion{Fe}{XVIII} intensities are for off-limb observations.
[1] \cite{curdt01}, [2] \cite{curdt04}.
}
\end{table}

\subsection{Observables provided by SPICE\label{S.observable}}

SPICE is capable of measuring the full spectrum in its two wavelength bands.
To optimise the science data return within the given telemetry budget, only the full profiles of the strong emission lines (marked in Table\,\ref{T.lines}) will be measured routinely. This will provide the intensities, Doppler shifts and widths of the lines, from which the non-thermal broadening can be determined.
The accuracy of the line shifts determined through centroiding will be of the order of 5\,km\,s$^{-1}$ at the longer wavelengths (depending on the signal-to-noise ratio).
There is an option of on-board summing of the line profiles and, separately, of the adjacent background. This will be particularly useful for the weaker lines observed with shorter exposure times, where the limited signal-to-noise ratio may prevent determining line shift and width, but the intensity summed across the line can be obtained and downloaded, while using very little of the allocated telemetry.  
Maps of line intensities, shifts, and widths will be provided as high-level data products.

In addition to these products that are directly deducible from the line profiles, further higher-level data products can be derived.
First and foremost is the possibility to investigate the elemental abundances, and in particular the separation of the elements according to the first ionisation potential (FIP).
Through the FIP effect, there is a preferential enhancement of elements of low FIP compared to those with high FIP, often called the FIP bias \citep[][]{Fludra99, 2000JGR...10527217V}. The enhancement depends on the source region, see a recent review of observations in \citep[][]{Del-Zanna:2018yu}.

The carefully selected lines from low-FIP elements (S, Si, Mg, Fe) and high-FIP elements (H, C, O, Ne) will allow maps of the FIP bias to be produced from SPICE data.
The broad temperature coverage from 10,000\,K to 10\,MK (see Table\,\ref{T.lines}) will be well suited to study the thermal structure of the solar atmosphere, all the way from the chromosphere to the corona, occasionally even including hot flare plasma around 10\,MK. Spatial maps of the emission measure can be computed in this temperature range. 
The spatial resolution of SPICE (along the slit) will be about 4{\arcsec}.
At perihelion (0.3\,AU), this resolution corresponds to 1.2{\arcsec} for an instrument observing from Earth orbit.
This is comparable to, or better than most previous EUV spectrographs.
The spatial resolution of Hinode/EIS \cite[][]{2007SoPh..243...19C} is about 2{\arcsec}, and SOHO/SUMER \cite[][]{1995SoPh..162..189W} provided a similar or slightly better resolution.
Only the most recent IRIS spectrometer \cite[][]{2014SoPh..289.2733D} provides a resolution of about 0.4{\arcsec} that is significantly better than SPICE.

In contrast to EIS and IRIS, SPICE provides a  much more comprehensive temperature coverage of the transition region, with lines spaced closely over a wide range of temperatures (see Table\,\ref{T.lines}).
EIS is mostly sensitive to hot plasma above 1\,MK, while IRIS, designed to observe the Sun's chromosphere and transition region, is mostly blind to the temperature range from 0.3\,MK to 8\,MK.
While SUMER could observe a temperature range wider than SPICE, it had to step through wavelengths to record spectral profiles, so that it needed considerable time to cover line profiles emitted over the full temperature range. In contrast, SPICE records all the lines simultaneously. This is a major advantage when studying the often very dynamic solar atmosphere.

In terms of temporal cadence, SPICE is comparable to previous EUV spectrometers. The brightest lines can be observed with exposure times of 1-5\,s, while 30-60\,s exposures are envisaged for comprehensive coverage of weaker lines.
This allows dynamic phenomena to be followed in a fashion comparable to previous instruments (although with restrictions set by the limited telemetry).
Considering its performance, SPICE will provide unique sets of data. 
SPICE will be able to take full advantage of the special vantage points close to the Sun and from high latitudes offered by Solar Orbiter during the extended mission, and it will always operate in concert with other remote-sensing and in-situ instruments.

\subsection{Science together with other Solar Orbiter instruments\label{S.all.instruments}} 

Solar Orbiter's vantage point of out-of-ecliptic latitudes will allow an unprecedented view of the poles. SPICE will carry out the first-ever out-of-ecliptic spectral observations of the solar polar regions. SPICE will provide maps of outflow velocities and identify the sources of the fast solar wind inside the polar coronal holes, connecting them to solar wind structures observed by in-situ instruments. Joint observations with the SWA/HIS sensor will allow the testing of models of the fast solar wind \citep{Fludra18}.

The magnetic fields near the poles are poorly known, and Solar Orbiter will provide major advances in this direction through the PHI instrument \citep{Solanki2019a}.
In this context, SPICE can provide the response of the upper atmosphere to the surface magnetic field in a globally open magnetic environment.
The investigation of the emission from plasma over a wide range of temperatures will show if the heating mechanisms in the magnetically closed quiet Sun at lower latitudes are comparable to those in the globally open coronal hole regions near the poles.

Another important goal of SPICE, together with the other remote-sensing instruments, is to understand small-scale heating events in the corona.
Here the line profiles from SPICE will provide the crucial information on the turbulent state of the plasma through the analysis of the non-thermal broadening.
Likewise, propagating waves that transport energy through the atmosphere reveal themselves through the spectral profiles.
For example, a non-compressible wave like an Alfv\'en wave, will not be directly visible in imaging observations, but will leave an imprint in spectral data.
The very good coverage in temperature will allow SPICE to determine the resulting thermal structure of the transition region (through the emission measure) with considerably improved resolution compared to previous imaging and spectroscopic observations.
In combination with the EUV imaging observations by EUI \citep{Rochus2019a} that will provide diagnostics of the spatial and temporal evolution, these SPICE observations will offer a new and comprehensive picture of the state of the plasma over a range of coronal structures. In particular, SPICE can carry out further studies of the coronal heating, through correlations of the transition region emission with the magnetograms in active regions \citep{Fludra2010}. Studies of the ubiquitous magnetoacoustic waves, for example, above sunspots \citep{Fludra2001} or in coronal holes \citep{Banerjee2011} will also be possible. 

Solar Orbiter's vantage point at high latitudes will also provide a new view at structures near the equator.
SPICE and EUI will have a novel view of, for example, loops connecting the plage regions near the preceding and trailing sunspot in an active region.
Such loops run mostly in the east-west direction and can only be viewed edge-on from Earth.

SPICE will play a crucial role in understanding the coupled system of the Sun and the inner heliosphere, the overarching science goal of Solar Orbiter.
Only a spectrometer can provide reliable tracers connecting measurements by in-situ instruments in the inner heliosphere to the near-surface regions of the Sun, observed remotely.

On the one hand, the Doppler maps provided by SPICE will provide information on the source region of the solar wind streams, for example in the vicinity of active regions or within coronal holes.
On the other hand, SPICE will provide maps of the FIP bias that might be a helpful tracer to identify the source of the solar wind.
In such investigations PHI will provide measurements of the underlying (changing) magnetic field and EUI will provide images of the temporal and spatial structures of the chromospheric features through the Ly-$\alpha$ line.
Together, this opens new possibilities to study the acceleration and heating in the actual source region of a solar wind stream that will then be captured and characterised in terms of magnetic field, waves, and particle properties by the in-situ instruments: MAG \citep{Horbury2019a}, RPW \citep{Maksimovic2019a}, SWA \citep{Owen2019a}.

Eruptive events, such as CMEs, will disrupt the coronal structures and can be eventually detected by the in-situ instrument suite. The radiation from the associated flare will also be detected by STIX \citep{Krucker2019a}.
In those cases where the source region has been observed by SPICE, the spectrometer will provide crucial information on the initial stages of the magnetic disturbance.
The shocks and increased turbulence associated with such an event can be studied through line shifts, and widths and intensity enhancements of spectral lines. With SPICE, these can be followed closely as a function of temperature.
Such observations are essential to understanding how (and where) shocks form in the corona.
Ultimately, this is the key to understanding the generation of solar energetic particles which, again, can be directly measured by the in-situ instruments, in particular EPD \citep{Rodriguez2019a}, and which leave their trace through the FIP bias in the data acquired by SPICE.

\section{Instrument overview}
\label{sect-instr}
The SPICE instrument is an imaging spectrograph that records high resolution EUV spectra of the Sun. The SPICE optical design was first presented in \citet{spice:2013_SPIE}. The instrument optics consists of a single-mirror telescope (off-axis paraboloid operating at near-normal incidence), feeding an imaging spectrometer. The spectrometer also uses just one optical element, a Toroidal Variable Line Space (TVLS) grating \citep{Thomas2003SPIE}, which images the entrance slit from the telescope focal plane onto a pair of detector arrays. Each detector consists of a photocathode coated micro-channel plate (MCP) image intensifier, coupled to an active pixel sensor (APS). Particular features of the instrument needed due to the proximity to the Sun include: use of a dichroic coating on the telescope mirror to transmit and thus reject the majority of the solar spectrum (this overcomes the large heat load close to the Sun), a particle deflector to protect the optics from the solar wind, and use of data compression due to telemetry limitations. The mechanical design and layout of the SPICE Optics Unit (SOU) are shown in Fig.~\ref{fig:spice_sou}, and the optical path is plotted in Fig.~\ref{fig:optical_path}.

As shown in Fig.~\ref{fig:optical_path}, the light enters the instrument through the entrance aperture. Then an image is formed at the slit by the off-axis parabola mirror. The slit defines the portion of the solar image that is allowed to pass onto a concave TVLS grating, which disperses, magnifies, and re-images incident radiation onto two detectors. The two wavebands cover the same one-dimensional spatial field, and are recorded simultaneously. Details of the optical path are further described in Sect.~\ref{sect-opt}. 
The instrument contains four mechanisms:
\begin{itemize}
\item The SPICE Door Mechanism (SDM), which can be actuated to provide a contamination tight seal of the entrance aperture during non-operational periods (both during ground handling and non-operational periods in flight).
\item The telescope mirror is mounted onto a two-axis mechanism (tilt and focus), the Scan-Focus Mechanism, that is used to direct different portions of the solar image onto the selected entrance slit and to focus the telescope relative to the entrance slit. The image of the Sun is repeatedly scanned across the entrance slit. During each scan the image of the Sun is stepped across the entrance slit in increments equal to the selected slit width, such that the region of interest is completely sampled.
\item A Slit Change Mechanism (SCM) provides four interchangeable slits of different widths, one of which can be selected depending upon the science activities to be conducted. These slits have a 2\arcsec, 4\arcsec, 6\arcsec, and 30\arcsec\ width on the external field of view. They are interchangeable via a slit change mechanism and are arranged on the mechanism in this order.

\item A vacuum door mechanism on the Detector Assembly (DA). The MCP and image intensifier used to translate the incident EUV photons into visible light photons that can be detected by the detectors must be maintained either at vacuum or in zero humidity during ground handling. Therefore the detector assembly contains a door mechanism which is only opened during vacuum testing on ground, and opened finally once on-orbit.
\end{itemize} 

The instrument structure consists of a stable optics bench with Aluminium honeycomb core and Titanium inserts. Baseplate facesheets, side walls and lids are made of Carbon Fibre Reinforced Plastic (CFRP). This is isostatically mounted to the spacecraft panel by means of Titanium flexures. The structure is designed to have approximately zero coefficient of thermal expansion (CTE), therefore maintaining instrument alignment throughout the wide operating temperature range.

The instrument control function will be provided by a dedicated electronics box, the SPICE Electronics Box (SEB). The SEB drives and monitors all mechanisms, the acquisition and processing of all housekeeping telemetry and the processing and packetisation of science data. It controls and communicates with the detector front-end electronics (FEE) via a SpaceWire link. The SEB also contains the SPICE flight software (FSW), which is responsible for all control and monitoring of the instrument, plus the processing and compression of the science data to allow the data rate and volume requirements to be achieved.

\section{Optical design}
\label{sect-opt}
\begin{figure*}
\includegraphics[width=2\columnwidth]{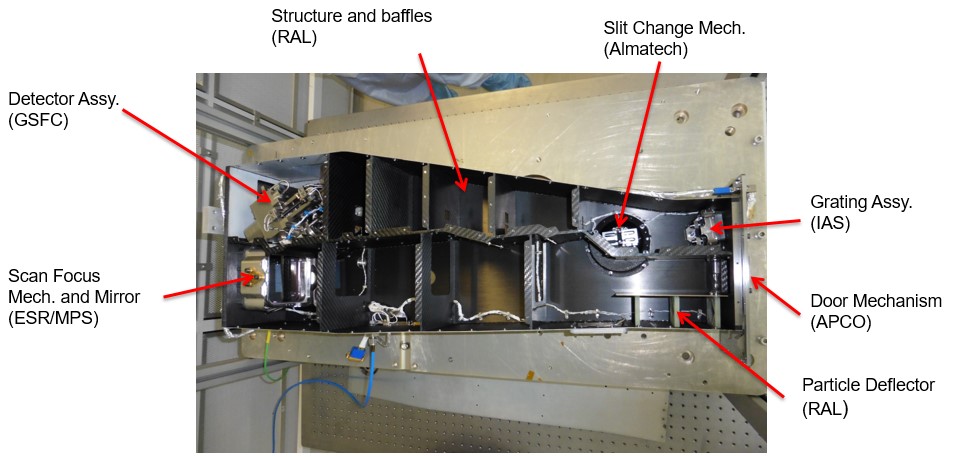}
\caption{Optics Unit of the SPICE instrument. In this top view into the SPICE Optics Unit, its key components are identified, along with the providing institutes and companies.}
\label{fig:spice_sou}
\end{figure*}

\begin{figure}
\includegraphics[width=\columnwidth]{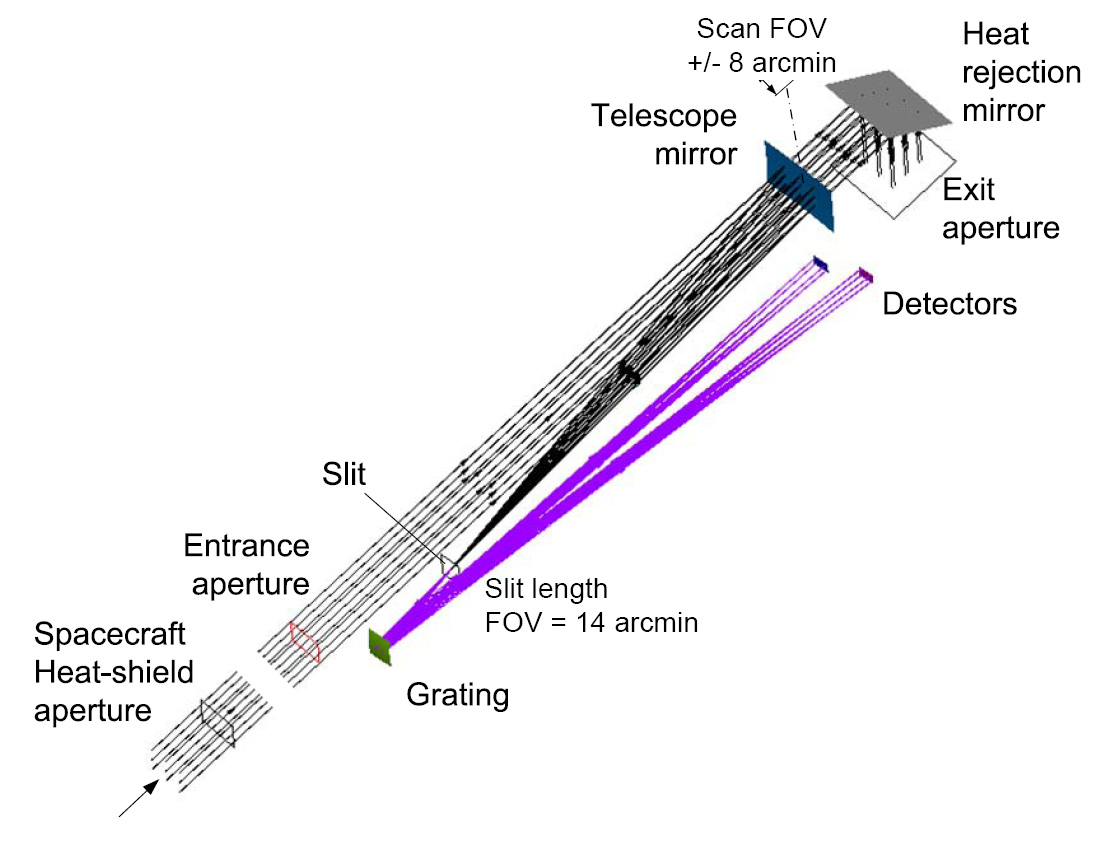}
\caption{SPICE optical layout. The system parameters are listed in Table~\ref{table:opticalParams}.}
\label{fig:optical_path}
\end{figure}

SPICE is a grating-spectrometer (Figs.~\ref{fig:spice_sou}, \ref{fig:optical_path}) where a portion of the solar disc is imaged by the single-mirror telescope onto the spectrometer entrance slit. The physical length of the slit is \textasciitilde 2\,mm, and for the mirror focal length (see Table~\ref{table:opticalParams}) it gives  an angular size of 11\arcmin\ (size of the along-slit instantaneous field of view on the Sun, oriented solar north-south). The slit is imaged by the diffraction grating  on the two array detectors. The grating has a concave toroidal surface form, which images the slit with magnification as given in the table; the physical size of the slit image at the detectors is $\sim$11\,mm. This spectrometer magnification also scales the telescope focal length to give a system effective focal length of \textasciitilde 3.3\,m. This is needed to give the required imaging-spectrometer spatial and spectral sampling (\textasciitilde 1\arcsec/pixel and \textasciitilde 0.01\,nm/pixel), with the given physical pixel-spacing of the array detectors (see Table~\ref{table:opticalParams}).

\begin{figure}
\includegraphics[width=\columnwidth]{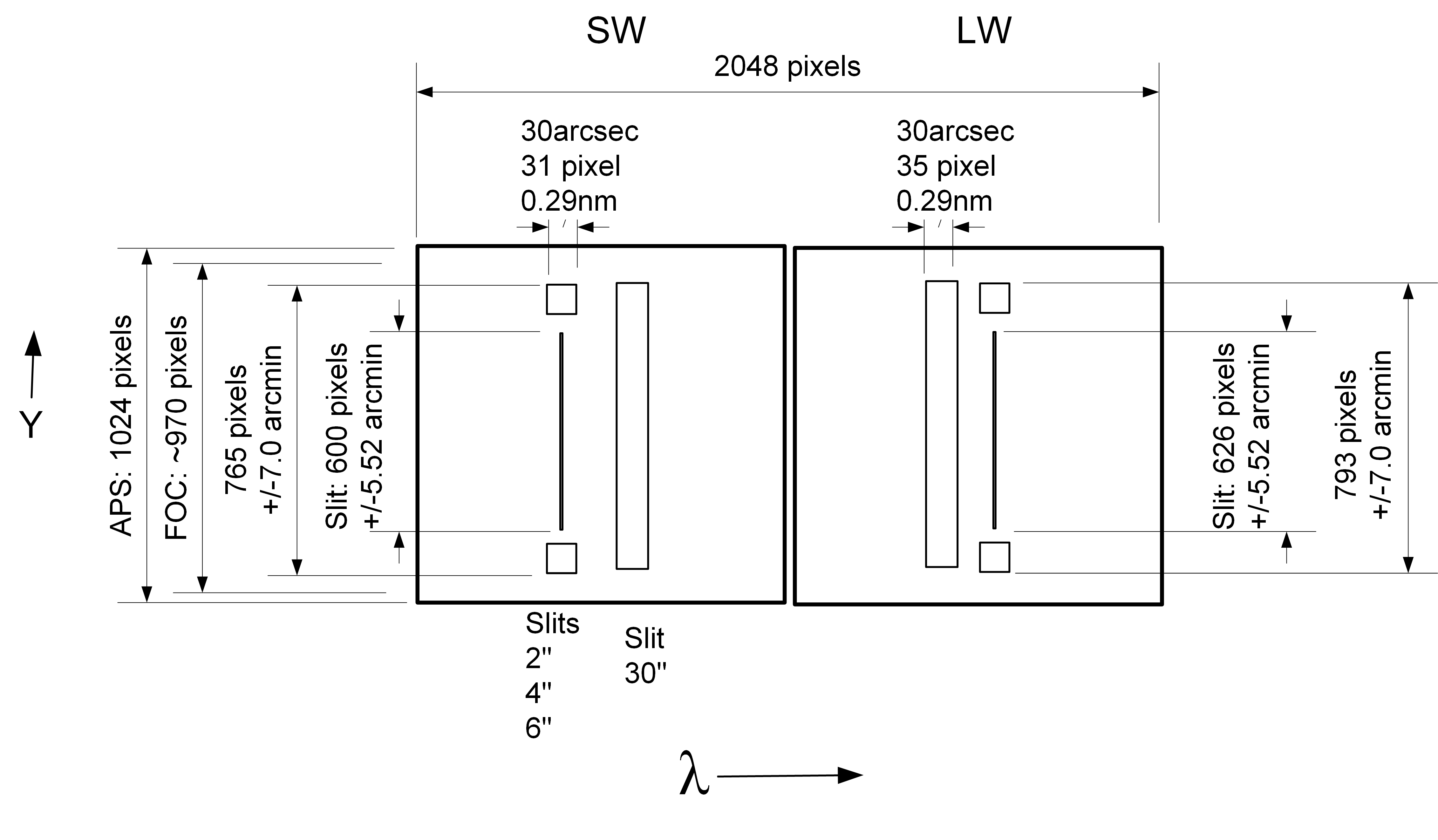}
\caption{Field-of-view diagram for SPICE, showing detector size, slit sizes, and spectral ranges (SW: short wavelengths, LW: long wavelengths, FOC: fibre optic coupler).}
\label{fig:FOV_diagram}
\end{figure}

The two detector arrays are separated in the focal plane in the dispersion direction according to the two chosen bands, short wavelengths (SW, 70.4\,nm--79.0\,nm) and long wavelengths (LW, 97.3\,nm--104.9\,nm). They are each based on a format of 1024$\times$1024\,pixels. The slits include square alignment apertures at each end (so-called dumbbells, shown in Figs.~\ref{fig:FOV_diagram} and \ref{fig:slit}), such that the total slit image length is \textasciitilde 14\arcmin. The detectors are oversized with respect to this image size such that they provide images with spatial sampling along-slit of approximately 14\arcmin/800\,pixels, or \textasciitilde 1\arcsec/pixel. In the spectral direction the sampling is \textasciitilde 9\,nm/1024\,pixels, or \textasciitilde 0.009\,nm/pixel (cf.\ Table \ref{table:opticalParams}). 
The dispersion relation of the spectrometer is given by the grating equation:
\begin{equation}
\sin(\theta_m) = m\cdot\frac{\lambda}d + \sin(\theta_i)\, ,
\end{equation}

\noindent
where $d$ is the ruling spacing, $m$ is diffraction order, $\theta_i$ is the angle of incidence and $\theta_m$ the angle of diffraction, with values as given in Table \ref{table:opticalParams}. 

In order to create the 2D images, the 1D instantaneous-FOV (slit FOV) is scanned laterally, over a range of 16\arcmin, by rotation of the telescope mirror. The mirror is mounted on a flexure rotation mechanism, driven by a continuous range mechanism (piezo actuator), but with a functional minimum step size of 2\arcsec, due to the mechanism encoder and control system used. The full range of the spectral imaging is thus an area of sun of 14\arcmin$\times$16\arcmin\, times 2$\times$9\,nm wavelength, sampled at \textasciitilde 800$\times$480 points spatially, times 2048 spectral points (maximum size of $x$-$y$-$\lambda$ data cube). 
The scan is controlled by closed loop, with a minimum time per step of 0.25\,s for a step size of less than 1 arcminute.  This gives a best maximum frame rate of 3\,Hz. For fixed-scan observations, the overheads are reduced to 0.1s per frame,  leading to a maximum frame rate of 5\,Hz (for the minimum exposure time of 0.1s). 

\begin{table}
\caption{Optical system parameters of the SPICE instrument.}
\begin{tabulary}{\columnwidth}{LL}
	\hline\hline
	{\textbf{Parameter}} & \textbf{Value}\\
    	\hline
  {\textit{Telescope}} & \\
    Entrance aperture size & 43.5\,mm $\times$ 43.5\,mm\\
    Distance entrance aperture to mirror & 770.8\,mm\\
    Focal length of parent paraboloid & 622\,mm\\
    f-number & 14.30\\
    Plate scale at slit & 3.02\,\SI{}{\micro\metre}/\arcsec\\
    Instantaneous FOV & Slit length:  11\arcmin, plus $\pm$0.5\arcmin\,'dumbbell' apertures, oriented solar north-south\\
    Width, rastered FOV & 16\arcmin\\
	\hline
   {\textit{Slits}} & Widths: 2\arcsec, 4\arcsec, 6\arcsec, and 30\arcsec\\
	\hline
  {\textit{Spectrometer}} &\\
    Wavelength range, SW & 70.387--79.019\,nm (1st order)\\
    Wavelength range, LW & 97.254--104.925\,nm (1st order)\\
    & 48--53\,nm (2nd order)\\
    Slit to grating distance & 128\,mm\\
    Grating groove density and $\pm$ chirp & 2400 $\pm$ 1.7\%\,mm$^{-1}$\\
    Grating image distance, SW mid-band & 692.43\,mm\\
    Grating image distance, LW mid-band & 720\,mm\\
    Spectrometer magnification (image-distance / slit-distance) & \textasciitilde 5.5\\
    Grating ruling period \textit{d} & 1/2400\,mm\\
    Grating angle-of-incidence & -1.7584$^\circ$\\
    Grating angle-of-diffraction (at SW, LW) & +8.5498$^\circ$ at 74.7\,nm +12.2398$^\circ$ at 101.12\,nm\\
    Dispersion, at image plane & 0.0095  at 74\,nm 0.0083 at 101\,nm (per pixel spacing)\\
	\hline
    \multicolumn{1}{l}{\textit{Detector}} & \\
    Pixel spacing & 0.018\,mm\\
    Angle-of-incidence on detector (SW, LW) & 27.33$^\circ$ at 74.7\,nm,  35.33$^\circ$ at 101.12\,nm\\
    \hline
   {\textit{System focal length (SW)}} & 3.371\,m\\
    \hline
  {\textit{System spatial plate scale}} & 1.101\arcsec/px at 74\,nm, 1.059\arcsec/px at 101\,nm\\
    \hline
\end{tabulary}
\label{table:opticalParams}
\end{table}

\subsection{Imaging resolution}
The narrowest slit is 2\arcsec\ wide.  Because the spectrometer magnification is the same in spatial and spectral directions, this corresponds to 0.02\,nm spectral width geometrically. However, when the optical-system and detector resolution effects are added, the net spectral resolution is 0.04\,nm (full-width half-maximum (FWHM) of line-spread function (LSF): \textasciitilde 4\,pixels). The contributors to imaging resolution are listed in Table~\ref{table:imagingResolution}, and each of the contributions is explained in the relevant subsections below.

\begin{table}
\caption{Imaging resolution contributions.}
\begin{tabulary}{\columnwidth}{LC}
	\hline\hline
    & \textbf{Contribution in pixels (spatial and spectral directions)}\\
    \hline
    Design, including nominal aberration, 2\arcsec\ slit-width $\times$ 2-pixel binning & 2\\
	\hline
    Optical Component tolerances contributions & 2.5\\
    \hline
    Position tolerances (build plus in-flight) & 1.5\\
    \hline
    Detector PSF (FWHM) & 2\\
    \hline
    Residual spacecraft jitter (10\,secs) & 1\\
    \hline
    \textbf{Total (RSS)} & \textbf{4.2}\\
    \hline
\end{tabulary}
\label{table:imagingResolution}
\end{table}

\subsection{The telescope mirror}
SPICE has a single-mirror telescope. The mirror is an off-axis paraboloid made of UV-grade fused silica substrate with a clear aperture of 95\,mm $\times$ 95\,mm and focal length of 622\,mm. The central area of 50\,mm $\times$ 50\,mm on the substrate has a thin reflective coating of boron carbide (B$_4$C). This single-layer coating is a novel design, which is a result of a compromise providing 30\% EUV reflectance while transmitting the rest of the solar spectrum (UV/VIS/IR) to space via a 45$^\circ$ fold mirror and an exit aperture. Such a dichroic design greatly reduces the heat load inside the instrument. The thickness of 10\,nm of the boron carbide coating was found to be advantageous for this purpose \citep{Schuehle:2007}.

The mirror substrate and the exit aperture are oversized with respect to the reflective aperture surface. The size is required to pass the angular range of the whole solar disc, for Solar Orbiter pointing at any part of it. The rear side of the substrate has an anti-reflective coating to maximise transmission of the solar spectrum passing through and beyond the reflective front coating. 

For imaging in the EUV, the mirror surface quality must be high. Thus, the figure error was specified as \textasciitilde $\lambda$/20 RMS. Measured at 633\,nm the RMS figure deviation of the flight mirror was 0.028 waves.
Also, the surface roughness must be low to limit scattered light from the whole solar disc while only a very small part of it is passing on to the spectrograph. The micro-roughness was specified as $<$ 0.2\,nm RMS, and the roughness measured by atomic force microscopy was 0.17\,nm.

At perihelion, the solar flux at the mirror is \textasciitilde 13 times the solar constant, and despite the low absorptance of the mirror over most of the solar spectrum, this results in approximately 3\,W of absorbed power, leading to a centre-to-edge thermal gradient of \textasciitilde 20$^\circ$C in the mirror, as the silica has low thermal conductivity. However it also has low thermal expansion, so the resulting `swelling' of the mirror's  front surface is small (predicted \textasciitilde 0.04\,\SI{}{\micro\metre}), which is not significant for aberrations, and is included in the term 'optical component tolerances' in table 3. This was also verified by tests, using a mirror illuminated with a solar-simulator UV beam, while monitoring its surface form using an optical test interferometer. 

The mirror is mounted on a scan-focus mechanism (SFM, see Sect.~\ref{sect-da-sfm}), which performs the angle scan (see above) using a piezo drive, as well as a motor-driven focus adjustment (range of $\pm$0.5\,mm). This focus adjustment is to allow for possible changes in the optics assembly dimensions from on-ground to in-flight, in particular for the varying thermal environment during the mission.  

Due to the proximity to the Sun the telescope mirror surface and the boron carbide coating are vulnerable to the high fluence of solar wind particles expected during the mission. As a protective measure, a solar wind particle deflector is included in the entrance baffle of the instrument structure. It consists of conductive plates with an applied voltage of $-2.5$\,kV, which will create an electric field strong enough to deflect incoming low-energy solar particles such as to prevent them from reaching the mirror.

\subsection{Spectrometer Slits}
\begin{figure}
\includegraphics[width=\columnwidth]{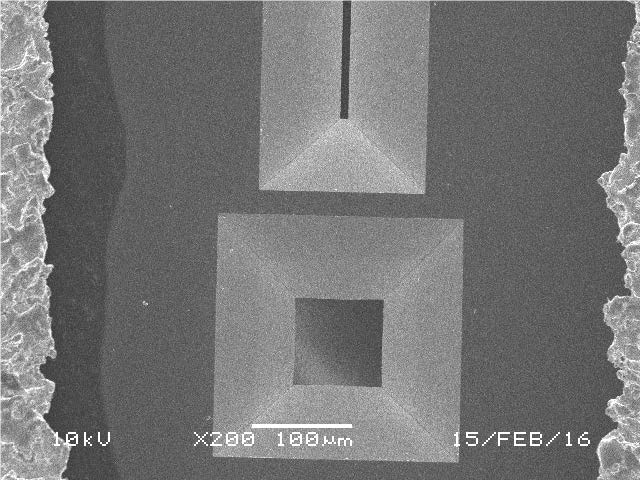}
\caption{Electron-microscope image of the rear of a SPICE slit (etched silicon). The end of a slit and one of its square dumbbells are shown.}
\label{fig:slit}
\end{figure}

The four slits are arranged in-line on a single frame, and to change slits this is raised and lowered within the telescope focal plane, on the linear slit change mechanism. Each slit is an aperture etched into a silicon slice of 0.5\,mm thickness (vee-groove etch, as shown in Fig.~\ref{fig:slit}), and gold-coated. The three narrow slits are 11\arcmin\ long (i.e.\ having 2\,mm physical length), but also with a small square aperture near each end (i.e.\ at a distance of $\pm7$\arcmin\ from centre). These serve to image small regions of the sun (0.5\arcmin $\times$ 0.5\arcmin) to obtain pointing information during the observations (see Fig.~\ref{fig:FOV_diagram}). An electron-microscope test image showing the end of one slit and its square aperture is shown in Fig.~\ref{fig:slit}.  The 30\arcsec\ slit (also known as a 'slot') is 14\arcmin\ long, and has no additional square apertures.  This element allows pseudo-spatial images to be acquired from isolated spectral lines, which can be used for certain science studies (e.g.\ movies), or for collecting instrument calibration data (e.g.\ detector flat-fielding).  

\subsection{Diffraction grating}
The diffraction grating is of TVLS type, as developed for this type of solar spectrographs. This enables slit-to-array detector imaging to be performed directly with the grating, in order to be able to dispense with additional mirrors normally required for aberration-control (chromatic astigmatism), and thereby greatly increasing the EUV throughput (for near-normal incidence the reflectivity of coatings at these wavelengths is only \textasciitilde30\%).
The aberration control over the range of the two detectors requires both toroidal form and varying line-spacing (`chirp'), that need to be precisely matched. The toroid radii of curvature have to match to \textasciitilde 1\%. The grating is holographic, 2400\,lines/mm (at grating centre), and with linear `chirp' variation of \textasciitilde 1\% of this across the used aperture. This level of linear variation is itself controlled to \textasciitilde 5\%  to match the toroid radii. The surface optical quality has to be similar to that of the telescope mirror, and the used aperture is much smaller (by the factor \textasciitilde 622\,mm/128\,mm). The reflective coating of the grating is of the same material as that of the mirror, boron carbide, but the thickness was increased to 20\,nm to increase reflectivity. This is possible due to the negligible solar heat load on the grating. 

The diffraction efficiency of the grating is a critical parameter, and was measured (using synchrotron radiation) to be \textasciitilde 9\%  (absolute efficiency).
For the spectrometer build tolerances, the large magnification is a challenge for the grating focus-setting and the correct alignment of the spectrum on the detector. For a de-focus blur radius equal to 1\,pixel, the axial distance of the grating from the slit has to be set to within $\pm$50 \SI{}{\micro\metre}. This is a practical challenge in the planning. Since the grating only works optically in the vacuum ultraviolet (VUV), the set up in air has to be done  to this accuracy by dead-reckoning, meaning by mechanical metrology using reference surfaces on the grating substrate and the slit mount. The alignment method was to then use a VUV test immediately after this set-up (i.e.\ before completing the build), to confirm this critical focus and alignment of the grating. 

\subsection{Detectors}
 
The detector assembly is described in detail in Sect.~\ref{sect-da_design}.  The incident UV light is converted to visible light inside the assembly, for detection by two independent sensor arrays, each sized at 1024\,pixels square.  The active area is large enough to record images of the full length of the slits plus dumbbells, with some margin.  However, the detector area limits the wavelength range in each band (see Sect.~\ref{sect-perf}). The pixel pitch of 18\,\SI{}{\micro\metre} also sets the spectral and spatial sampling, which in both cases is over-sampled relative to the instrument resolution (see Tables \ref{table:opticalParams}, \ref{table:imagingResolution}).  

\subsection{Stray-light design}
There are two main effects, both due to non-ideal light-scattering in the optical system: out-of-field light and out-of-band light.  The out-of-field effect is the light from the surrounding scene, meaning outside of the FOV, that is scattered into the FOV. It has a worst-case for the viewing of the relatively faint corona (when spacecraft points at the limb, SPICE will view at up to \textasciitilde 8\arcmin\ above the limb). The relatively bright light of the solar disc scatters at instrument baffles and at the optical surfaces. This effect is mainly restricted to the telescope because most of the solar disc is blocked at the slit. The scatter from baffles (vane edges) is kept low by designing the vanes as oversized from the used FOV (i.e.\ the beam envelope defined by combination of the entrance aperture and the slit) and giving them sharp edges. Also the baffle material is CFRP which is absorbing to VUV, and the vanes are designed to block any grazing-incidence light paths from the structure surrounding the optics. The mirror roughness and particulate contamination are kept as low as possible ($<$0.2\,nm RMS, and $<$100\,ppm surface area, respectively).   

The out-of-band light is the diffuse scatter within the spectrometer that adds a constant background level to the measured spectra, adding to the photon noise. This effect is kept low, again by use of baffle vanes, particularly around the slit mechanism where there are surfaces close to the beam, by the roughness quality of the grating surface and its lines ruling (grating grooves) plus its cleanliness, and by the visible-light blocking of the detector photo-cathode (so called solar-blindness). The grating's final roughness after the etching of its grooves (\textasciitilde 40\,nm depth) is 0.8\,nm RMS. In the final instrument testing, when imaging spectral lines, the out-of-band level was found to be \textasciitilde 0.1\% relative to line peak, at \textasciitilde 0.1\,nm from line centre.

\section{Mechanical and thermal design}
\label{sect-mech}
\subsection{Mechanical design}
\begin{figure}
\includegraphics[width=\columnwidth]{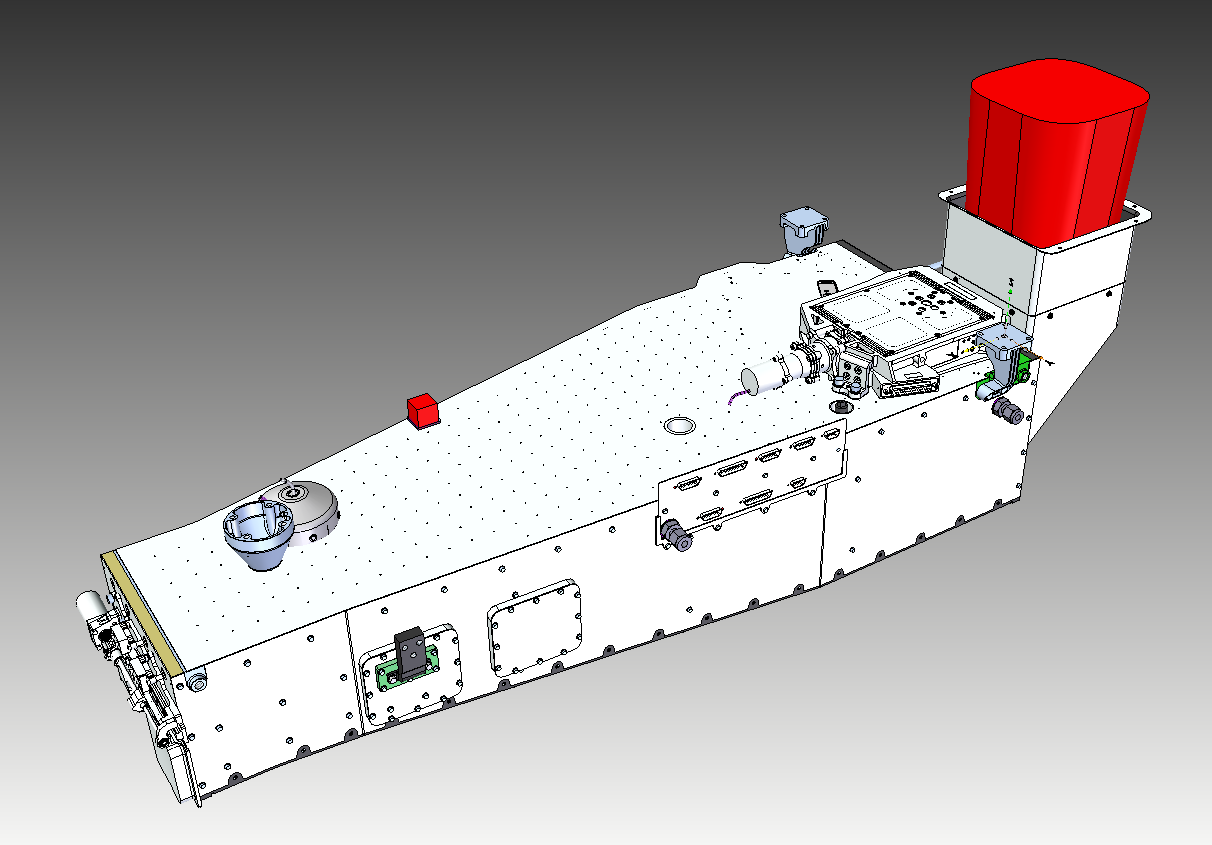}
\caption{SPICE Optics Unit: Details of interfaces to spacecraft including flux-exit aperture (upper right).}
\label{fig:sou_interfaces}
\end{figure}

The SPICE Optics Unit (SOU) is primarily made of a CFRP-and-aluminium honeycomb optical bench structure, onto which most of the subsystems are mounted, along with CFRP panels to produce a light-tight enclosure and stray-light baffles. The rear of the SOU houses the heat rejection mirror and baffle that allows the unwanted infrared (IR) radiation from the entrance aperture to pass out the rear of the instrument and out to space. These key interfaces can be seen in Fig.~\ref{fig:sou_interfaces}.
The SOU has a total mass of approximately 13\,kg and maximum dimensions of 1100 $\times$ 350 $\times$ 280\,mm.

\begin{figure}
\includegraphics[width=0.348\linewidth]{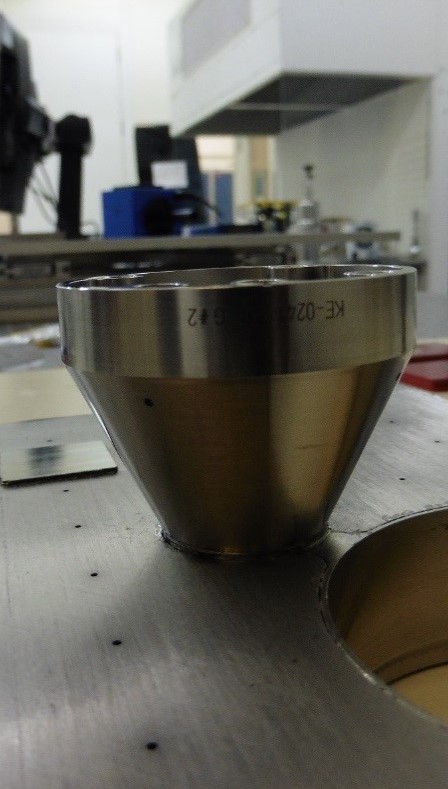}
\includegraphics[width=0.652\linewidth]{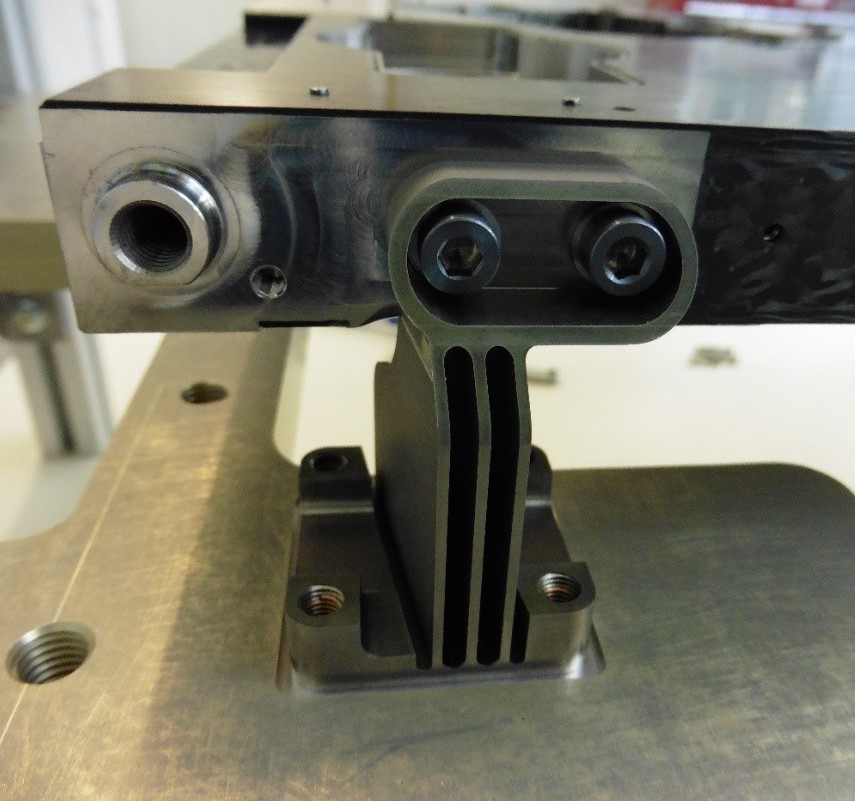}
\caption{SPICE Optics Unit: Interface Mounts (left: fixed mount, right: bladed flexible mount).}
\label{fig:sou_mounts}
\end{figure}

The SOU interfaces to the spacecraft panel with three `quasi-kinematic' mounts; one fixed foot at the front of the unit and two mounted on blades at the rear that allow for the differential CTE between the optics bench and the spacecraft panel -- these are shown in Fig.~\ref{fig:sou_mounts}. The mounts (manufactured from titanium alloy Ti-6Al-4V) are optimised in order to be compliant enough to compensate for the \textasciitilde 1\,mm in-plane difference in CTE between the spacecraft and optics bench (without distorting the bench), while also being stiff enough for the SOU to meet the minimum resonance frequency requirement of 140 Hz and survive the launch loads (the actual resonance frequency is 224 Hz). The blade mounts were manufactured in the RAL Space Precision Development Facility and heat treated and surface treated in order to maximise their performance.

\begin{figure}
\includegraphics[width=\linewidth]{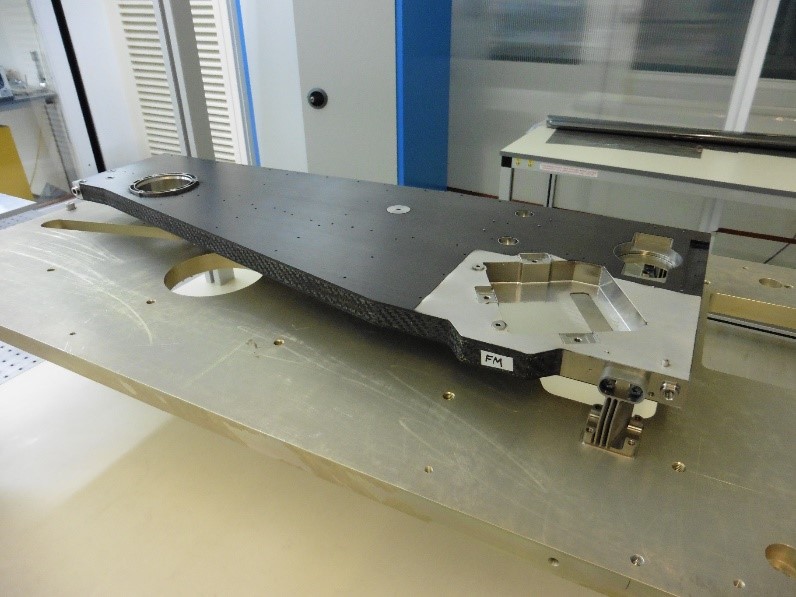}
\includegraphics[width=\linewidth]{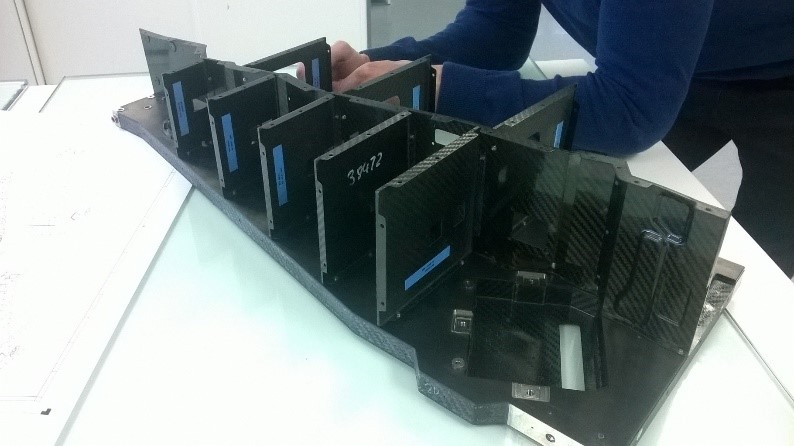}
\caption{SPICE optical bench structure without (top) and with (bottom) internal stray-light baffling.}
\label{fig:optical_bench}
\end{figure}

The optical bench prior to installation of the subsystems can be seen in the upper panel of Fig.~\ref{fig:optical_bench} and the single-skin honeycomb panels used as stray-light control and stiffening ribs can be seen in the lower panel. The design is based on the use of a very low-CTE CFRP in order to maximise the thermo-mechanical stability of the instrument across a wide range of temperatures. Initial coupon testing using interferometry demonstrated CTE of 0.2--0.7\,ppm/\textdegree C for a representative sample, although the final optical bench measurement showed a higher ($2.5\pm0.5$\,ppm/\textdegree C) figure which could be accepted by the use of margin within the alignment budget. This increase is thought to be due to stronger interaction between the CFRP face sheet, the adhesive film and the honeycomb core than had been assumed in the theoretical model.

\subsection{Thermal Design}
The main requirements of the thermal design of SPICE are:
\begin{itemize}
\item To manage the solar load and maintain all instrument components to within their operational and non-operational temperature limits
\item To control the detectors to a stable temperature of less than $-20$\,\textdegree C during all operational periods
\item To minimise the heat flow that is rejected (either radiatively or conductively) to the spacecraft thermal interfaces
\item To ensure that the primary mirror is warmer than its surroundings during the cold early phases of the mission, to avoid contamination of its surface
\end{itemize}

The primary thermal challenge for the SOU is managing the extreme heat input (\textasciitilde 17\,kW/m$^2$) during operation at perihelion. The thermal control system must also be compliant during periods with little   solar loading, with the conditions during the Earth and Venus gravity assist manoeuvres required for orbit adjustment (with additional planetary IR and albedo thermal loads) and with conditions with the instrument off-pointed from its central axis. 

The SOU is accommodated within the spacecraft, behind its heat shield.  A feedthrough in the heat shield provides a view for the instrument. The heat shield includes a door that can be used to prevent direct sunlight entering the instrument feedthrough during non-operational periods. 

The SOU is, with the exception of two designated radiative and conductive heat rejection interfaces, thermally decoupled from the spacecraft. Conductive decoupling is achieved through the use of low thermal conductivity titanium for the kinematic mounts, and the natural isolation required by the quasi-kinematic mount design. Radiative decoupling is achieved through the application of a low-emissivity aluminised coating to the external surfaces of the SOU structure.    

The thermal design utilises a synthetic quartz (Suprasil\textsuperscript{\tiny\textregistered} 300) primary mirror with a 10\,nm thick B$_4$C coating that reflects the EUV radiation of interest for science but transmits the visible and near-infrared solar radiation with little absorption. Consequently, much of the high-flux solar radiation entering through the aperture passes through the instrument and is then reflected to space by the heat-rejection mirror (HRM) attached to the rear of the instrument. The HRM assembly is a CFRP structure mounted to the rear of the instrument that houses a highly reflective diamond turned aluminium fold mirror.  

The majority of the solar radiation that is reflected within the SOU by the primary mirror is intercepted by three pre-slit heat rejection mirrors (mounted before the slit) and reflected to a single high-absorptivity heat dump radiator. As these heat loads are relatively low, this re-radiates to the internal surfaces of the spacecraft. The pre-slit mirrors are configured so that just the required science beam is passed through to the slit (anything that reflects on to them via the primary mirror is not part of the science beam). Baffles also intercept radiation that either diverges as it comes into the instrument or is off-axis due to the spacecraft pointing away from the Sun centre. 

At perihelion, the instrument can survive thermally when the spacecraft is off-pointed by up to 3.5$^\circ$ in any axis in steady-state, and by 6.5$^\circ$ for a period of up to 50\,seconds. This gives sufficient margin for nominal spacecraft operations which will never point beyond the solar limb, maximising at 0.94$^\circ$ at  perihelion.  At  the  distance  of  0.7\,AU,  the  steady-state  off-pointing limit increases to 4.5$^\circ$, and at 0.95\,AU the spacecraft can be oriented at any attitude without thermally affecting SPICE. 
The above thermal analysis shows how far of an off-axis pointing SPICE can survive (without necessarily being required or able to do science). 

The primary thermal design driver is to manage, at perihelion, the solar load incident through the 52$\times$52\,mm aperture in the spacecraft heat shield. Fig.~\ref{fig:sou_interfaces} illustrates the thermal model predictions for the nominal perihelion case, with end-of-life (EOL) thermo-optical properties. About 66\% of the 31.7\,W entering the SOU cavity is transmitted through the primary mirror and reflected directly to space by the HRM. The remaining 10.7\,W is absorbed within the SOU structure. Of this, 2.5\,W is directed to the heat dump radiator, where it is radiated to the spacecraft. In the perihelion case, the surrounding spacecraft temperature is specified as 50\textdegree C. The instrument structure averages about 55\textdegree C due to the addition of absorbed solar loads and internal dissipation. It is noted that the view of the HRM structure to deep space provides radiative cooling, which reduces the heating effect of the absorbed solar loads. About half of the internal absorption of solar flux occurs at the primary mirror (at the coating and within the silica substrate). The primary mirror is therefore warmer than its surroundings, operating at about 70\textdegree C.

The detector assembly is conductively and radiatively isolated from the instrument surroundings at the mounting interface using polyether ether ketone (PEEK) standoffs. Heat generated internally and the low levels of parasitic heat to the assembly are rejected to the spacecraft-provided cold element interface. This allows the active pixel sensors within the DA to be passively cooled. They are then individually PID (Proportional-Integral-Differential) controlled by SEB-powered heaters to a set-point of $-$20\textdegree C.

\begin{figure}
\includegraphics[width=\linewidth]{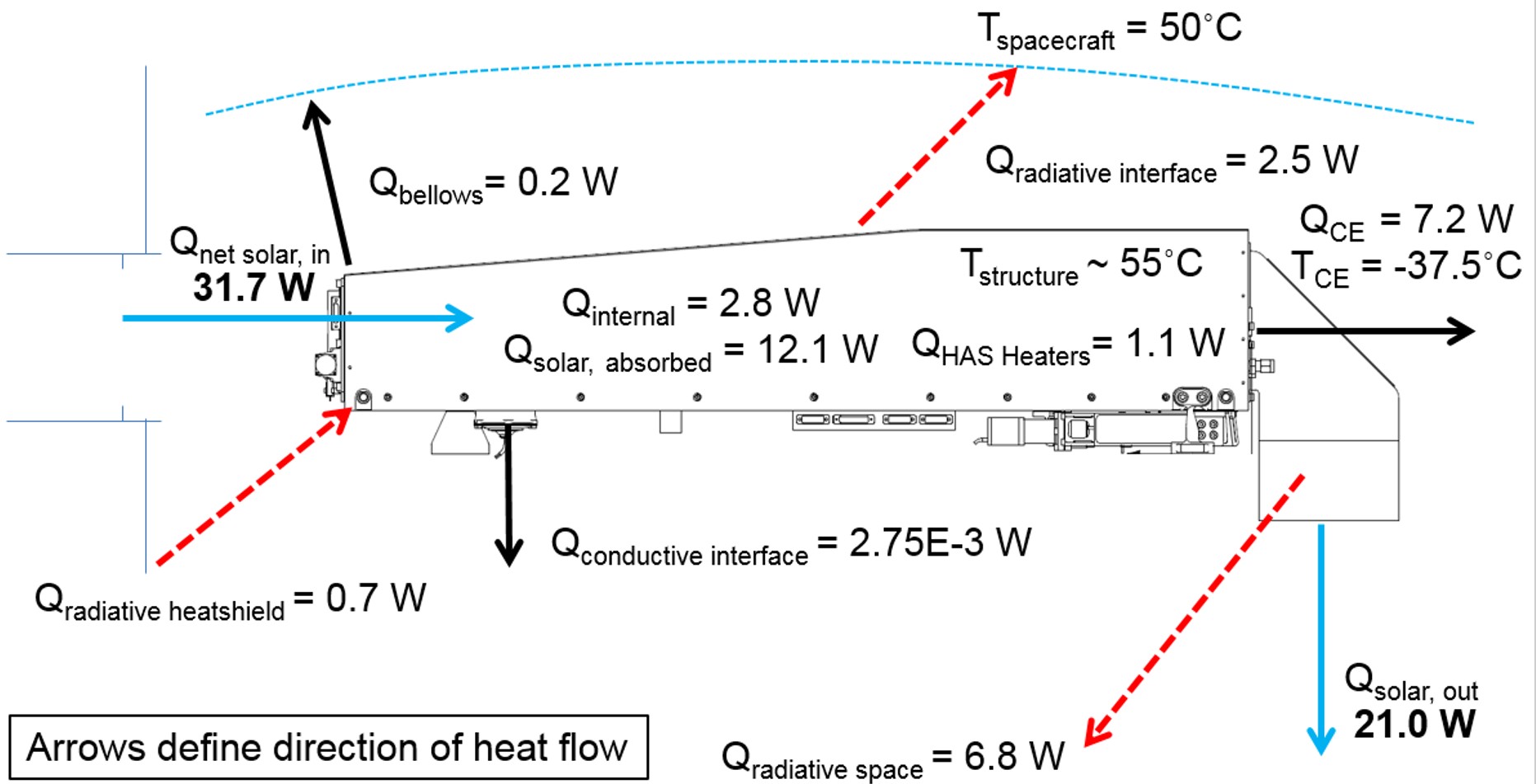}
\caption{Predicted thermal energy balance for the SPICE Optics Unit during perihelion operation, at end-of-life.}
\label{fig:thermal_energy}
\end{figure}

\section{Mechanisms and detector assembly design}
\label{sect-da}
\subsection{SPICE Door Mechanism}
The SPICE Door Mechanism (SDM) provides a contamination seal at the entrance aperture of the instrument. It protects the highly sensitive internal optics during non-operational periods during the ground integration and test phase, and during the cruise and non-operational phases in flight. The mechanism consists of the door itself (with a highly reflective finish and spherical shape to reject the incoming high intensity solar flux during some flight phases), which is articulated on linear bearings and driven by a stepper motor (with reduction gear-box) and ball screw. The door maximum temperature is \textasciitilde 125$^\circ$C when closed at perihelion. However, the door and mechanism design is not thermally qualified for the transient case of being opened or closed while sun-illuminated, and this means that for these operations, the outer heat-shield door must be closed.

\begin{figure*}
\includegraphics[width=0.5\textwidth]{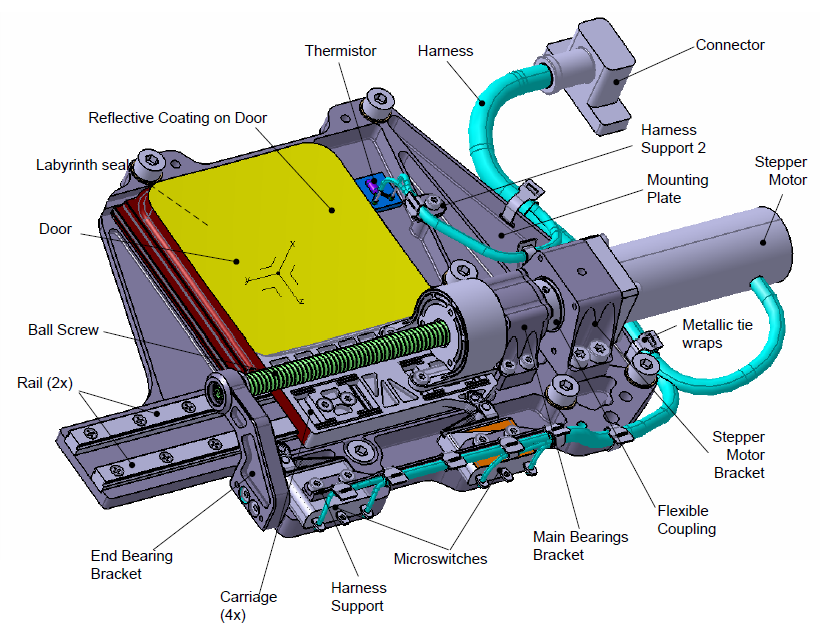}
\includegraphics[width=0.5\textwidth]{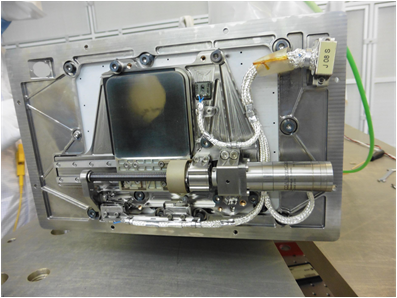}
\caption{Door Mechanism. The drawing on the left highlights its components, the picture on the right shows the assembled flight model.}
\label{fig:spice_door}
\end{figure*}

The SDM provides the defining aperture for the instrument optical design, including a knife edge to control the stray-light impact of the aperture and reject the oversized beam passing through the heat shield feedthroughs. The door forms a labyrinth seal which is contamination-tight, but allows purging of the optical cavity during Assembly, Integration and Verification (AIV) up to launch. The SDM is designed and qualified for up to 100 open-and-close cycles so that it can be used repeatedly both during AIV and during flight between the remote-sensing windows (to limit contamination from the spacecraft entering the instrument). The door opening and closing operations involve driving through motor steps, between the end positions. These are detected by position switches at each end of the range; in addition, the steps are counted. At the step rate used, the time required to open and close is approximately 50\,s. The component parts of the SDM design are illustrated in the left panel of Fig.~\ref{fig:spice_door} and the flight model mechanism (integrated to the front panel of the SOU) is shown in the right panel.

\subsection{Scan-Focus Mechanism}
\label{sect-da-sfm}
The required range of motion for the mirror is $0-8$\,\arcmin\ in rotation (about an axis parallel to the slit-direction), and $\pm 0.5$\,mm in linear motion (in the focus direction). At the same time, the mechanism must have high stiffness in the other degrees of freedom to maintain alignment and stability. This is achieved by using a flexure-based design, with multiple blade flexures for strength. It is a two-stage design comprising (1) a linear stage for focus, driven by a roller-screw mechanism with stepper motor, and (2) a rotation (scan) stage. This is mounted on the linear stage, and is driven by a piezo-electric actuator via a lever arm. The mirror assembly itself is mounted on this rotation stage, and the adjustments for optical alignment during the build are made at this interface. 

The motions of both stages are sensed by linear variable differential transducer (LVDT) sensors. The sensor for rotation is connected to a lever arm, which amplifies its displacement to allow a suitable accuracy of the sensor for the control system. This rotation (scan) has closed-loop control to give the required stability and step resolution ($\approx$\,2\,\arcsec). This mirror scanning is used regularly during many of SPICE's observing sequences, and the response time for scan stepping with a small range is typically 0.25\,secs. The linear (focus) stage has step size 0.4\,mm, and position accuracy is $<2$\,\SI{}{\micro\metre} at a given temperature. During observations the focus setting is changed only infrequently, for example during calibrations or after change in instrument temperature (depending on mission phases).

\begin{figure*}
\includegraphics[width=\textwidth]{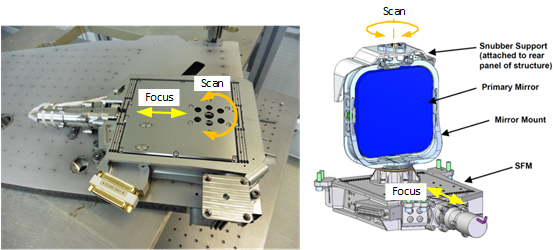}
\caption{Scan-Focus Mechanism. The picture on the left highlights focus and scan directions, the drawing on the right shows how the instrument's primary mirror is mounted on the mechanism.}
\label{fig:scan_focus}
\end{figure*}

\subsection{Slit Change Mechanism}

The slit change mechanism has the function of positioning any one of the four slits into the active slit position, to the required absolute and stability tolerances (in particular as regards the spectrometer focus and spectral calibration). The slits are mounted in-line in a single carrier mounted to the mechanism stage. The physical length of each slit is approximately 2.5\,mm (slit length including dumbbells, for 14\arcmin\ angular size on the Sun), and the spacings between adjacent slits are 5, 6, and 5\,mm, so the total range of motion needed is 16\,mm plus margin. To change from one slit to any other, is a single linear movement. Due to the fixed mechanism speed the time taken is a minimum for move between adjacent slits, to a maximum for a move between the end slits

For the mechanism design the slits carrier is mounted between two large leaf-spring flexure blades (titanium), in order to provide the required motion range (along-slit direction) while maintaining stiffness in the across-slit (spectral) and focus axes. It is driven by a stepper motor, which drives a Rollvis satellite screw, which translates the rotary output of the stepper motor into linear motion. The step size of the slit motion is 0.02\,mm. The assembly includes mu-metal magnetic shielding that greatly reduces the magnetic signature of the Sagem stepper motor. 
A limit switch is implemented at one end of the range of travel, 0.25\,mm from the 6\arcsec\ slit.  All operations of the mechanism are performed by counting the number of motor steps moved relative to this limit switch, which defines the `zero' position.  Standard operations will include driving the mechanism back to the limit switch on a regular basis (after every few slit changes), to ensure that the positional reference is maintained.  The standard travel speed of 0.5\,mm/s allows an adjacent slit to be selected in $10-12$\,s.  A slower drive speed is used for finding the limit switch, but the reset operation can always be completed in less than 1\,minute, from any given starting position.  

\begin{figure*}
\includegraphics[width=\textwidth]{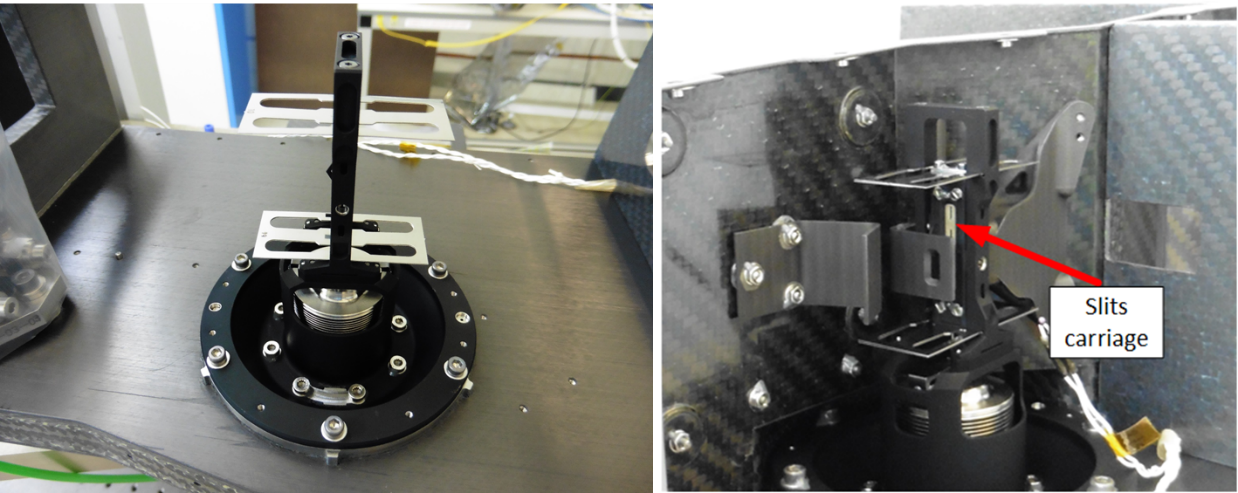}
\caption{Slit Change Mechanism.}
\label{fig:slit_change}
\end{figure*}

\subsection{Detector Assembly Door Mechanism}

The primary purpose of the detector assembly door mechanism is to control contamination, mainly molecular, as well as humidity, which can both be detrimental to the accuracy of the detector as well as the functionality of the MCP.  It is critical that the door never be opened when not at vacuum or in <30\% humidity environment, or the intensifiers will be damaged.  

The actuator for the door mechanism is a controllable drive actuator. This motor was selected as a smaller mass option to the other motors on the SPICE project while still meeting requirements.  The motor is able to achieve 2.3\,Nm of torque across the operating temperature range of the SPICE DA. Conversion of the rotary stepper motor motion into linear displacement will be performed using a worm gear, while a pin-and-slot feature on the door part will allow the door to open and close.

Due to the contamination requirements of the instrument and the relatively undemanding life requirements, dry lubrication is used on the motor mechanism. The door actuation cycles will be low for this mechanism.  The qualified number of cycles for this mechanism is one in flight and 20 during ground testing.

\subsection{Detector Assembly design}
\label{sect-da_design}
\begin{figure*}
\includegraphics[width=0.48\textwidth]{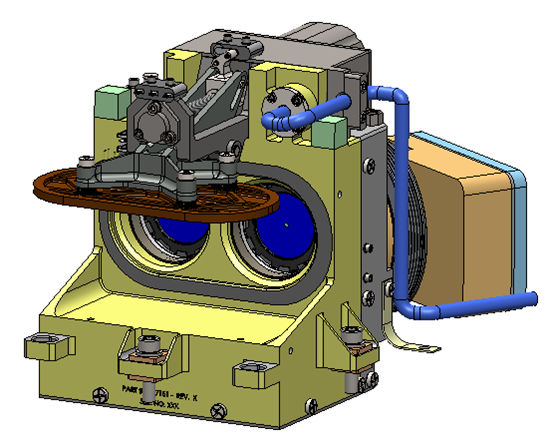}
\includegraphics[width=0.52\textwidth]{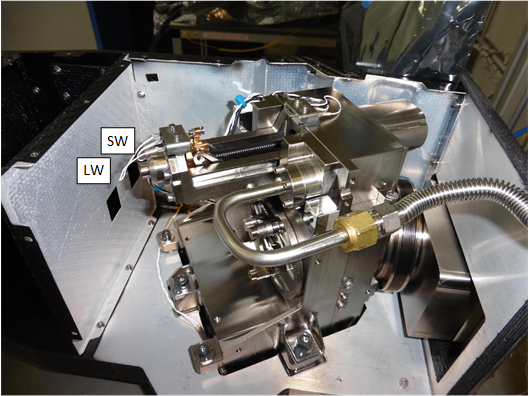}
\caption{Left: Detector Assembly design concept (with door open), Right: Detector Assembly within the SPICE SOU (door closed to protect the MCPs).}
\label{fig:da_design}
\end{figure*}

The SPICE DA (Fig.~\ref{fig:da_design}) consist of two independent, identical, intensified APS camera systems mounted in a common sealed housing. Each camera consists of a HAS2 (High Accuracy Startracker 2) 1024$\times$1024\,pixels format complementary metal-oxide semiconductor (CMOS) APS with digital readout electronics fed by a KBr-coated micro-channel plate intensifier.  This DA is radiation-hard, and has significant heritage through multiple orbital and sub-orbital missions.

\begin{figure}
\includegraphics[width=\linewidth]{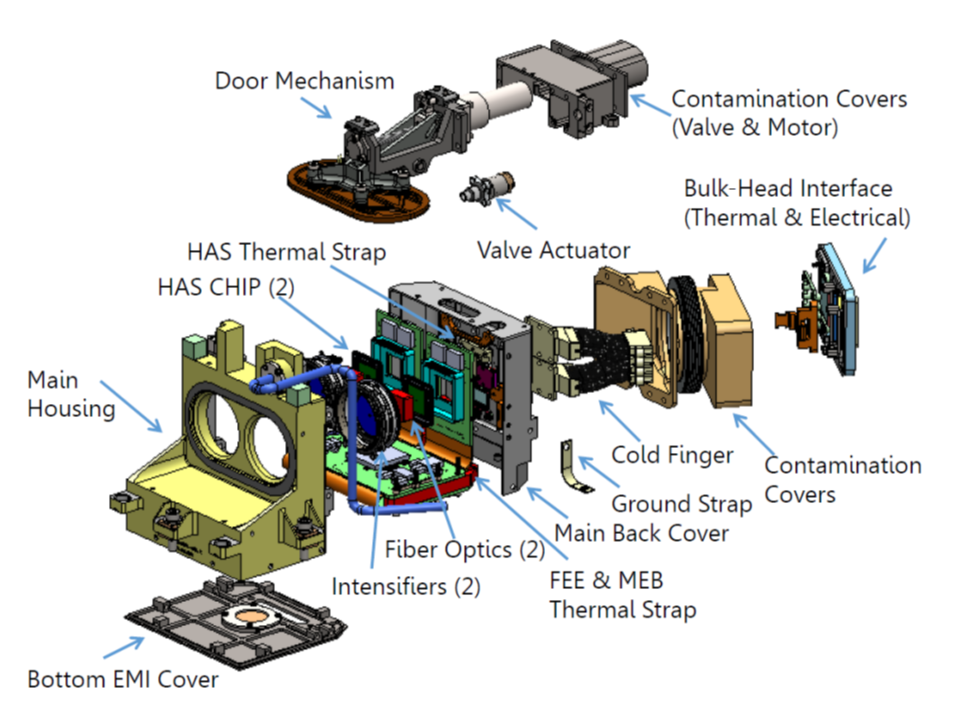}
\caption{Detector Assembly, exploded view.}
\label{fig:da_exploded}
\end{figure}

The main elements of the DA are shown in Fig.~\ref{fig:da_exploded}.  The SPICE intensifier tubes consist of a micro-channel plate and a phosphor screen packaged in a lightweight housing. Each provides a nominally 25\,mm diameter active area that circumscribes the sensor active area. A KBr photocathode is deposited on the front surface of the MCP to enhance response in the SPICE passbands, while remaining visible-light blind. Photons absorbed by the KBr layer are converted to photoelectrons and amplified through the MCP based on the applied voltage across the MCP. The nominal MCP voltage from the high-voltage power supply (HVPS) is 850\,V (up to 1200\,V possible), unless adjustments are necessary due to detector ageing.  

Electrons exit the 6\,\SI{}{\micro\metre} MCP pores and are accelerated by a 2800\,V potential across a sealed proximity gap (0.5\,mm) onto an aluminised phosphor screen deposited onto a fibre-optic output window. Electrons are converted into a visible-light image at the phosphor screen. The resulting image is transferred through a fibre optic coupler to the APS sensor. A direct bond between the fibre optic and the APS seals the APS active area and eliminates environmental contamination on the APS. The fibre optic coupler is similarly bonded to the MCP fibre-optic output window, again eliminating environmental contamination.  The MCPs are scrubbed to stabilise the MCP gain against localised charge depletion. After scrubbing, the MCP housing is maintained at low humidity levels until launch to maintain sensitivity.

The APS detectors are cooled (by conductive link to the spacecraft cold-element) to minimise the dark-current, and the temperature is stabilised using heaters on the detector thermal straps to achieve $-20^\circ$C.  This approach allows the spacecraft cold element temperature to vary during the orbit, while still achieving stable detector performance (which is sensitive to temperature).  

The front-end electronics (Sect.~\ref{sect-elec}) are located close to the focal plane to maintain the integrity of the clocking and analogue output signals. Within the front-end electronics, a Field-Programmable Gate Array (FPGA) will accept configuration commands and generate the timing signals needed to operate the HAS2 sensor as configured. The analogue video signal from the sensor will be digitised to 14 bits precision.

\section{Electronics (including FPA hardware and flight software)}
\label{sect-elec}
The SPICE instrument has two electronic units: the SPICE Electronics Box (SEB), which contains the control and data processing, and the Front-End Electronics (FEE), which form part of the detector assembly within the SOU.

\subsection{SPICE Electronics Box}
The SEB circuit boards are:
\begin{enumerate}
\item High Voltage Power Supply (HVPS) for the MCP, Gap and Particle Deflector
\item Data Processing Module (DPM) incorporating the spacecraft interface, signal processing and instrument control
\item Two Mechanism Interface Modules (MIM) driving the motors, scan mirror, position sensors and SOU heaters 
\item Low Voltage Power Supply (LVPS) which converts the spacecraft 28V to the SEB internal supplies
\item Backplane which is used to transfer the various signals between the other 4 circuit boards.
\end{enumerate}

\subsection{Mechanical Design}
The SEB utilises a single chassis with 4\,mm wall thickness that houses the six circuit assemblies. The five plug-in cards  are designed with wedge type card retainers that interface to machined card slots, providing both structural support and a conductive thermal path to the chassis. After all modules have been inserted into the chassis, a single front panel is installed on the front, which along with rabbet joints at all panel to panel mating, minimises electromagnetic radiation exiting the chassis, and cosmic radiation entering the chassis. The chassis uses vent holes sized such that there is adequate ascent depressurisation capability while maintaining a small aspect ratio to give good EMC (electromagnetic compatibility) performance.

\subsection{Data Processing Module}
The DPM provides essential data processing, instrument commanding and science operations management for the SPICE instrument. The DPM is the central part of the SEB and performs the following functions:
\begin{itemize}
\item Hosts 8051 microcontroller and flight software (FSW)
\item Command \& telemetry interface to the spacecraft
\item Control of instrument mechanisms
\item Command \& control of FEE
\item Image processing and compression
\item Control of the high and low voltage power supplies
\end{itemize}
Control and management of the mechanism operation is provided by the DPM through a combination of software functions operating in the 8051 and hardware resources within the command and control FPGA. The DPM mechanism control interfaces include:
\begin{itemize}
\item SPICE Door Mechanism
\begin{itemize}
\item Stepper motor for opening and closing SPICE door
\item Thermistor for monitoring mechanism temperature
\item Microswitches for detecting when door position
\end{itemize}
\item Slit Change Mechanism
\begin{itemize}
\item Stepper motor for moving the slits carriage, movable in half or full steps
\item Microswitch for detecting when slit is in home position
\item Thermistors for monitoring the temperature of the slit mechanism
\end{itemize}
\item Scan Mechanism
\begin{itemize}
\item Lead zirconate titanate (PZT) actuator for controlling scan mechanism
\item Linear variable differential transducer (LVDT) sensor for monitoring the position of the scan mechanism  
\item Closed loop control algorithm
\item Thermistor for monitoring temperature of mirror mount
\item Thermistor for monitoring temperature of scan stage
\end{itemize}
\item Focus Mechanism
\begin{itemize}
\item Stepper motor for controlling focus mechanism
\item LVDT for monitoring focus mechanism position
\item Thermistor for monitoring temperature on focus stage
\end{itemize}
\item Detector Assembly
\begin{itemize}
\item Stepper motor for controlling detector assembly door.
\item Thermistors for monitoring the temperature of the detector assembly
\item Microswitches for monitoring detector door position
\end{itemize}
\item SPICE Optical Unit
\begin{itemize}
\item Thermistors for monitoring the temperature of the SPICE optical unit
\item Heaters
\end{itemize}
\end{itemize}
The DPM controls the high voltage power supply outputs through a set of digital to analogue converters located on the DPM and routed via the backplane to the HVPS together with a number of discrete digital signals for on/off control etc.

The DPM uses a single SpaceWire interface for FEE configuration and control. The FEE and science acquisition settings (e.g.\ exposure duration, pixel selection for readout, gain) are provided by the DPM resident FSW utilising a simple set of registers within the DPM image processing FPGA. These registers are used to pass configuration information via the SpaceWire interface to the FEE.

The LVPS module provides the power interface to the spacecraft.  It accepts the 28V input, provides noise filtering, and converts it to the required voltages to run the other boards and components inside the SEB (1.5, 2.5, 3.3, 5.0, 5.6, and $\pm12$\,V).  A filtered 28\,V supply is also provided for the heaters and stepper motor drivers.  The design considers the need for good EMC performance, in order to minimise any conducted emissions that could affect other instruments on the spacecraft. 

\subsection{High Voltage Power Supply}
The HVPS for the SPICE instrument provides the high voltage (HV) required to run two MCP and intensifier pairs as well as a particle deflector. The HVPS consists of five individual supplies sharing a common board, each with its own set of control signals within a common low interface, and each responsible for a HV output.

The high voltages are:
\begin{itemize}
\item Two MCP supplies: 0\,V to 1275\,V (nominal 850\,V) at up to 50\,\SI{}{\micro\ampere} for each supply. 
\item Two gap supplies: 0\,V to 3570\,V above MCP voltage at up to 10\,\SI{}{\micro\ampere} for each supply.
\item Particle deflector supply: -2500\,V at 10\,\SI{}{\micro\ampere}. 
\end{itemize}

\subsection{Flight Software}
\label{sec:FSW}

The SPICE FSW is written primarily in the C programming language, with a small amount of assembly language for initial boot up at power on. The Keil Integrated Development Environment (IDE) was used for creating, editing, compiling, linking and testing (through 8051 emulation) the FSW.

The SEB FSW runs on a Microsemi Core8051 IP core that is part of the DPM CPU FPGA. The primary function of the FSW is to act as the data manager for the SPICE instrument by sending and receiving messages over the SpaceWire command and telemetry interface to the spacecraft, controlling the mechanisms, heaters and HVPS outputs, managing the acquisition and processing of science data, monitoring and reporting instrument health, and helping manage the safety of SPICE. 

The FSW resides in two non-volatile memories of the DPM:
The boot code image is stored in the 32\,KB PROM, and the science code image is stored in the 256\,KB EEPROM device. 
The EEPROM contains two science code images (prime and redundant) that can be up to 64\,KB and the lookup tables (LUTs), similarly stored as primary and redundant images, each up to 64\,KB in size.

\begin{figure}
\includegraphics[width=\linewidth]{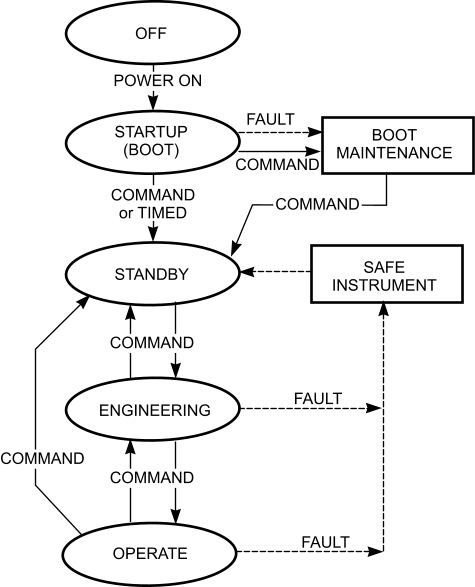}
\caption{Software Modes.}
\label{fig:software_modes}
\end{figure}

The modes for SPICE are Startup, Standby, Engineering and Operate as shown in Fig. \ref{fig:software_modes}.
When SPICE is switched on the FSW immediately enters STARTUP mode and runs the boot code contained in the PROM to execute initialisation and self-test routines. The boot image uses the SpaceWire interface to the spacecraft to implement functions such as memory load and dump so that the LUTs or the science code images can be updated or checked. Prior to transition to the STANDBY mode, a check is performed on the science and LUT images stored in EEPROM to ensure that they are uncorrupted. If the images are sound, they are copied to SRAM and run automatically. 

When the boot to science image transfer occurs, the FSW will be in STANDBY mode which is a stable and safe configuration for the SPICE instrument. Transition to the ENGINEERING mode is accomplished with a command. This mode allows for the ramping of the HVPS outputs connected to the micro-channel plates (MCPs), gaps and particle deflector, prior to transitioning to OPERATE mode. A command is required to transition to OPERATE, which is the mode used for science image acquisition, processing and telemetering.

SPICE uses a bespoke, Consultative Committee for Space Data System (CCSDS)-based science packet format to organise the data efficiently, while still including enough metadata to make it self-describing.  This is required due to the complex structure of SPICE science observations, which are organised by windows (wavelength ranges) and compressed in one of several formats (see Sect.~\ref{sec:compression}).  Detailed information is presented in the SPICE Data Interface Control Document (ICD). The science packet contains image header information in the first packet, followed by image data in the remaining packets. A checksum of the image header and data is appended and stored in the last packet.

If an anomaly occurs during Engineering or Operate mode, the fault detection, isolation and recovery system generates an event message and then has a subsequent action of either to continue operation if the anomaly is benign or to perform recovery activities.

\subsection{Front End Electronics}

\begin{figure}
\includegraphics[width=\linewidth]{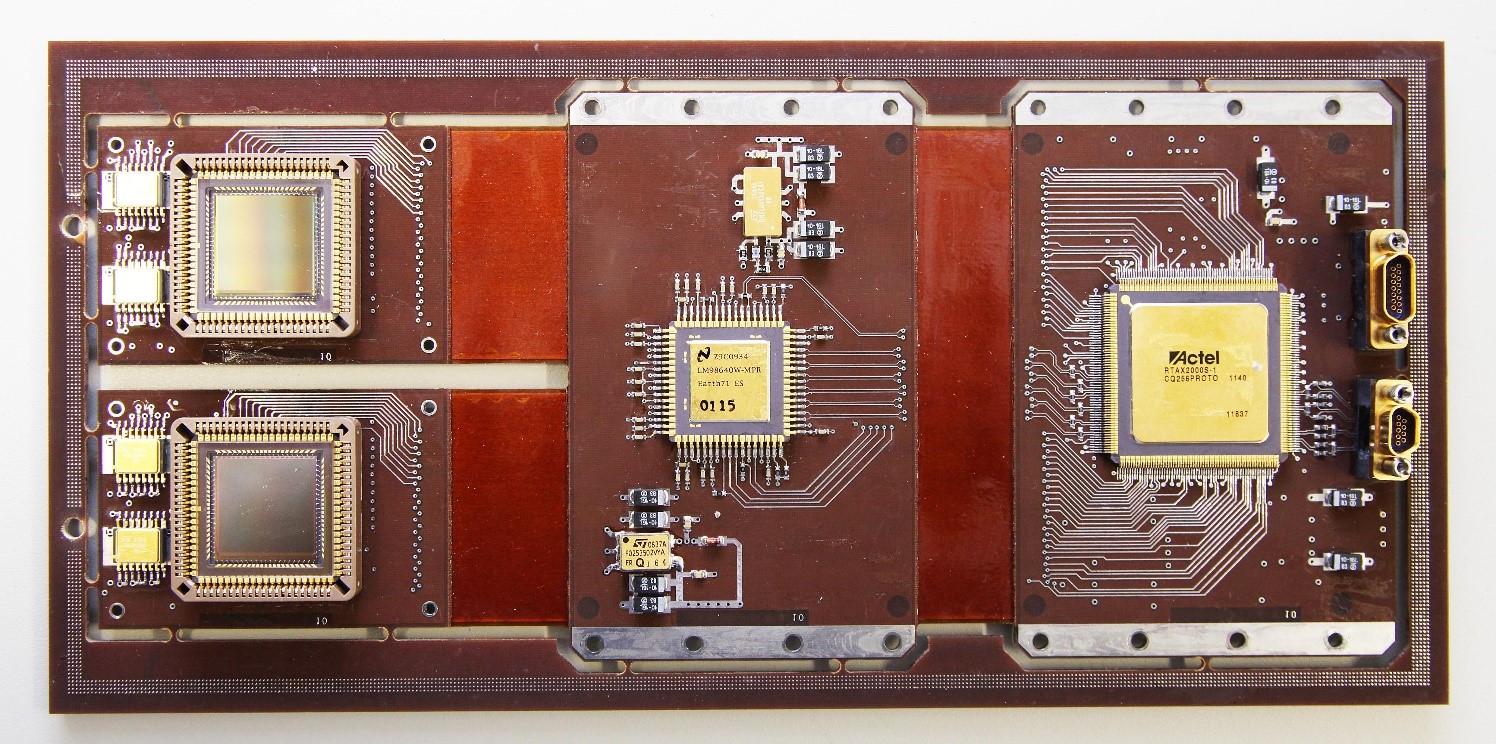}
\caption{Front End Electronics board before assembly into the Detector Assembly (left to right: boards 1 to 3).}
\label{fig:fee_board}
\end{figure}

The Front End Electronics (FEE) boards are contained within the Detector Assembly (see Sect.~\ref{sect-da_design})
and due to space constraints are a single assembly comprising three circuit boards with flexi-rigid interconnects.  This approach
uses less volume and improves the reliability of the FEE due to reduction in number of individual connectors.  The flexi-rigid 
approach also allows the boards to be `folded' up so that that HAS2 active pixel sensors can be mounted in the correct
position within the detector assembly.  The FEE is used to convert optical signals to electrical signals and transmit these to
the SPICE Electronics Box. The assembly contains two HAS2 active pixel sensors (board 1), an analogue to digital converter (board 2) and a control
FPGA (board 3), see Fig.~\ref{fig:fee_board}.

\begin{figure}
\includegraphics[width=\linewidth]{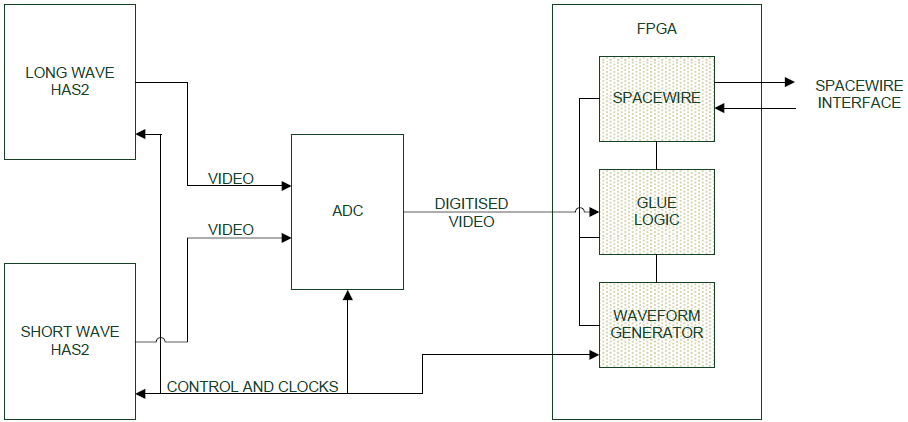}
\caption{Front End Electronics: Block diagram.}
\label{fig:fee_block}
\end{figure}
As shown in Fig. \ref{fig:fee_block}, the FEE provides the control and clock signals needed by the two HAS2 detectors, along with the readout links to a single ADC chip.  The HAS2 detectors are configured and operated in such a way as they appear to be one larger detector of 1024$\times$2048\,pixels; the 2048\,pixels being in the spectral direction and the 1024\,pixels being in the spatial direction.

Image size, readout sequences and window parameters are defined by programming the FEE's waveform generator and sequencer's internal readout table memory from the control and data acquisition interface. Exposures can be defined and timed externally to the FEE, but alternatively an internal timing routine can be set up within the waveform generator and sequencer if required. The FEE is highly configurable, allowing either full-frame or readout of up to 32 pre-defined regions of interest (windows). Windowed readout reduces the images size and thus bandwidth required for each image. 

The FEE is programmed by sending it two-byte SpaceWire packets which write data to the 8-bit registers in the FEE FPGA. In configuring the FEE for a particular mode of image acquisition, the SEB must:
\begin{itemize}
\item Load the required image readout waveforms and tables into the FEE waveform generator (WG) RAM 
\item Load the window registers in the FEE to select the required pixel columns (wavelength ranges) 
\item Configure the analogue front end, including the gains (coarse \& fine) and offset for each sensor 
\item Load configuration registers in the sensor.
\end{itemize}
The above RAM and registers are loaded by writing ASCII characters to a FPGA register. These characters represent the I2C data link `start' and `stop' signals as well as hex digits which convey addresses or data.

The FEE supports both destructive and non-destructive read out modes. Destructive readout is when the pixel is reset immediately after the signal level has been sampled. This black level reset signal is then subtracted from the signal level during pixel readout, thus eliminating any static pixel-to-pixel offsets of the sensor. Non-destructive readout mode is when the sensor is reset and then immediately readout. This black level image is then stored by the SEB. After the exposure time has elapsed another image is readout and again stored by the SEB. The corresponding black level image is then subtracted from the signal image. This is correlated double sampling, which eliminates static offsets as well as thermal (`kTC') noise.

The RAL SpaceWire Interface FPGA IP core provides the FEE's communications with the control and data acquisition interface. The communications interface is a SpaceWire adaptation of the IEEE1355 serial interface standard, featuring an LVDS-driven SpaceWire serial data link running at 100\,Mbits/s. The SpaceWire link can support a range of communication speeds, which are programmed by writing to registers. At reset, the transmit and receive links are configured to run at the default speed of 10\,Mbits/s.

\subsection{Science Concept of Operations \label{sec:studies}} 
A science `observation' campaign is expected to be a series of so-called `studies', which are scheduled in the mission timeline. A study consists of  a series of spectra (or sets of windowed spectral lines) acquired at successive positions of the primary mirror scanning mechanism. For the end user, a study can be thought of as one $\lambda, y, x$ raster of a $(\lambda, y, x, t)$ data set (for this purpose, sit-and-stare observations can be thought of as having an X step size of 0). A raster cube is commanded by sending a STUDY\_RUN telecommand with the number of rasters (study repetitions) to be performed as sole parameter. Studies are defined in on-board lookup tables (LUT), which are loaded with the required definitions for the operational period (see Sect.~\ref{sec:study-gen}). Some complex operations will also use macros, which are internal command sequences to be run on an execution engine within the FSW. Macros are also defined in the onboard LUT, and consist of a series of SPICE commands with relative time tags that determine when each command is executed. These can (and often will) include commands to run a study.  

When a study starts, the FSW reads the corresponding science LUTs which hold parameters to configure the SPICE instrument for science acquisition. The LUTs contain: slit choice; exposure time, scan mirror start, stop and increment (step size) information; the number of scan repetition `loops'; and window position, size, binning and processing information such as whether to compress the data (and the method of compression) or to keep the data in a raw state prior to packetisation. The FSW then commands the mechanisms, communicates with the Front End Electronics (FEE), sets up the Image Processing FPGA (IPF) for acquisition, configures the Discrete Wavelet Transform (DWT) Bit Plane Encoder (BPE) Application-Specific Integrated Circuits (ASICs) for compression (if desired), and coordinates the science data product packetisation process and transmission to the spacecraft mass memory unit.

The FEE must initially be programmed by the SEB to perform the desired acquisition of spectral or image data and then the Waveform Generator (WG) to execute the appropriate table to perform readout of the detectors. The resulting pixel values representing an image frame are then received by the SEB in the form of a SpaceWire packet from the FEE. The selection of pixels contained in the packet depends on how the WG and window registers have been programmed. The WG table that reads out the detector also determines the exposure by controlling the time between reset and readout of pixel rows. 

The WG table used for readout, and its component waveforms, also determines where SpaceWire end-of-packets (EOP) are inserted; the EOP is inserted after each complete (i.e.\ of all active windows) readout of the detector. For SPICE, the readout table is used to generate only a single acquisition (image) at a time.  The timing of the start of each acquisition is managed by the FSW, while the readout table controls the timing of all FEE operations within each acquisition.  This includes the necessary overheads for resetting the active pixels (i.e.\ the windows) within the image, which requires 0.42\,s for reading a whole array (i.e.\ 2048$\times$1024\,pixels).  The time to reset a smaller area depends on the number of wavelength pixels (i.e.\ how many of the 2048) used.  A typical science study includes up to 256 wavelength pixels (for eight windows of 32\,pixels wide), for which the reset overhead is just over 0.05\,s.  Normally this does not add any overhead to a study, as the scan mechanism takes longer than this to move to the next position, and the detector reset can occur concurrently. 

\subsection{Image Data Processing\label{sec:image_processing}}
The SPICE DPM IPF is a single FPGA containing the following functions: 
\begin{itemize}
\item One FEE SpaceWire Interface 
\item Black Level and Dark Current Corrections 
\item Binning in Y (spatial) and $\lambda$ (wavelength) 
\item Buffer Synchronous Dynamic Random Access Memory (SDRAM) Interfaces 
\item Flash Interface 
\item Fast Fourier Transform (FFT) Processing 
\item Interfaces to DWT and BPE ASICs Central to the science data acquisition and processing is a pipelined structure involving the following pipeline stages:
\begin{itemize}
\item Non-destructive and destructive pixel readout 
\item Selectable black level correction 
\item Selectable binning 
\item Selectable dumbbell extraction for alignment tracking
\item Frame accumulation for SHC (Spectral Hybrid Compression, see Sect.~\ref{sec:compression}).
\end{itemize}
\end{itemize}

The first processing blocks in the IPF after data is read from the FEE via the SpaceWire interface are the Corrections and Y Binning blocks. In the Corrections and Y Binning blocks, positive offset, black level subtraction, dark current correction, and Y binning can be applied to the incoming image data. These corrections are enabled based on FSW configuration data from LUTs loaded into the IPF. The processing steps are conducted in the following order with the result being stored in the window buffer:
\begin{enumerate}
\item Black level subtraction 
\item Positive offset 
\item Dark current correction 
\item Binning in the Y (spatial) direction (2, 4, 8, 16, 32, and 64\,pixels sizes). 
\end{enumerate}

FSW manages the contents of this dark current correction map by initialising its contents with data. The FSW also copies the data from the non-volatile storage for its calculations, scaling the pixel amplitude for the appropriate exposure time and writing the result in the window buffer for dark map subtraction use. This scaling module provides arbitrary multiplication of a 14-bit pixel value and 13-bit exposure value as well as left shift by 6, 9, 11, or 12 based on an input shift code. The exposure value and the shift code are provided to the IPF at run time by flight software. 

If Y binning is enabled and if any one (or more) of the pixels being binned are saturated above a specified 14-bit number ($\lid$16,383\,DN [Digital Number]) then the super-pixel which they are being summed into is set to a saturated 16-bit value (65,535 DN). After the applications of corrections, the data is transferred to the window buffer block as an intermediate storage location. The window buffer comprises a single SDRAM memory device. 

From the window buffer block, data is routed to the Lambda ($\lambda$) Binning block. Binning in the wavelength direction, if enabled, can be selected in 2, 4, 8, and 16 pixel sizes. Like Y binning, if any one (or more) of the pixels being binned are saturated above a specified 14-bit number ($\leq$16,383 DN) then the super-pixel which they are being summed into is set to a saturated 16-bit value (65,535 DN). After traversing the Lambda Binning block, pixel data enters the Fast Fourier Transform (FFT) Processing block. 

In the FFT processing block, if enabled, Fourier component calculation is performed on a per window basis. The FFT is computed across all Y rows within a given Y by $\lambda$ window. This is a 32-point FFT with 18 bits of precision. Data is sent to the FFT using either the lower 12 or 14 bits of unsigned data, pre-scaled to occupy the most-significant bits of the input. Data internal to the FFT is scaled at each of the internal FFT stages to prevent data overflow. Data retrieved from the FFT consists of the low 16 bits of output. There are 2 real and 15 complex coefficients per FFT. This amounts to 32 scalars, each being assigned to a particular `coefficient plane.' Note, if SHC is enabled, the FFT function is executed automatically as part of that process and precludes any use of lambda binning. 

After the FFT block, data enters the scan buffer block, which is a second SDRAM intermediate storage location. Additionally, alignment windows (also known as dumbbells) are filtered directly to the scan buffer after being received from the FEE if an observation was configured to collect them. Alignment windows are not corrected, binned, or FFT-ed. 

\subsection{Compression  \label{sec:compression}}

The scan buffer block is capable of organising multiple exposure frames of data accumulating a `cube' of X by Y by $\lambda$ images. From the scan buffer, data is transferred to the compression Interface block.

In order to meet the downlink telemetry volume constraints, the DPM incorporates a novel compression algorithm known as Spectral Hybrid Compression (SHC). SHC performs compression on the Fourier coefficients to achieve an up to 20:1 compression ratio. The data flow associated with the SHC algorithm includes compression in the X by Y plane for each FFT coefficient plane with the following options: 
\begin{itemize}
\item A menu of 8 possible configuration `recipes' that can be chosen from for best SHC compression are made available. 
\item Each recipe has different BPE settings. 
\item Compression recipes designate values for the configuration settings made available by the ASIC interface module, described below. 
\item Compression algorithm is CCSDS 122.0-B-1. 
\end{itemize}
Compression is performed by two specially programmed ASICs which operate at a maximum rate of 5\,MPixels/sec to reduce overall power consumption. Each ASIC is connected to two dedicated SRAMs for use during compression. The compression hardware works on individual slices of the `cube' consisting of several FEE images, corresponding to different scan mirror (i.e.\ X) positions. The `data cube' thus has two spatial axes and one wavelength axis ($x$, $y$ and $\lambda$). It is also possible to disable the compression methods such that raw pixel data is transmitted to the ground. 

The ASIC interface module within the IPF provides a mechanism to configure each ASIC independently with the image compression parameters required to achieve the desired compression function for the current image data. The DWT ASIC must be reset prior to reconfiguration (e.g.\ a change to the image height or width), and as a consequence all ongoing data must be flushed from the chip prior to the change. The BPE ASIC can be reconfigured while compression processing is ongoing and provides synchronisation mechanisms so new parameters are applied only on an image boundary. 

The ASIC interface module provides the following configuration settings: 
\begin{itemize}
\item {\tt Height}: Height of the image in pixels (32 to 1024). 
\item {\tt Width}: Width of the image in pixels (32 or 64). 
\item {\tt SegByteLimit}: This is the number of 16-bit compressed words used to represent the image. 
\item {\tt S}: The number of 8$\times$8 blocks of pixels in the image. 
\item {\tt SignedPixels}: Whether the input image data is signed or unsigned. 
\end{itemize}
The other ASIC configuration settings have been left at their default values so that the compression conforms to the CCSDS 122.0-B-1 specification. 
After compression, data is transferred to the output buffer block where it is stored until transmission to the spacecraft using the appropriate science packet structure (see Sect.~\ref{sec:FSW}). This buffer is used for both compressed and uncompressed science data.  

\section{SPICE testing}
\label{sect-env}
In this section we describe the testing methods used to verify the key performance and engineering requirements of the instrument. The molecular and particulates cleanliness of the instrument is a key requirement, and so in the instrument build and throughout all of the testing the cleanliness control is an important aspect.
The overall flow of the tests described in this section are summarised in the flow diagram in Fig.~\ref{fig:inst_test_flow}.

\subsection{Cleanliness control}
In common with all space instruments that observe in the UV spectrum, and especially in the EUV, SPICE is highly susceptible to degradation effects of contaminants:
\begin{itemize}
\item Darkening of optical surfaces, meaning loss of reflectivity, due to build-up of deposited molecular contamination
\item Increase in scattered light (stray-light effects) due to particulate contamination on optical surfaces.
\end{itemize}

For both of these, the effect is so great that very stringent control of these contaminants is needed throughout the flight-hardware program. The parts in the optics unit must be manufactured as initially very clean or `pristine', and then protected from contamination (and verified as clean), throughout the ground testing. For the molecular effect, the main source is hydrocarbons and silicones such as from wire insulation and epoxies in the assemblies. The total allowance is 200\,ng/cm$^2$ per optical surface (\textasciitilde 50\,nm film thickness), for which the light-throughput loss at the SPICE bands is \textasciitilde 20\%. For the scattered light effect, the allowance is 100\,ppm, for which the effect is then as low as the scatter from optical surface roughness ($<$0.3\,nm RMS).

To achieve these levels a contamination-control plan had to be made, for all of the build and test procedures to adhere to, which included the following:
\begin{itemize}
\item In the design, selection of approved materials with correct mechanical and low-outgassing properties, and preparation procedures;
\item In the fabrication of all parts, use of  precision cleaning, followed by pre-bake at the highest possible temperature before assembly;
\item For use of epoxies, use of low-outgassing types, verified by batch-testing, de-gassing before use, and curing using buffer gas flushing to prevent re-deposition of outgassing products.
\end{itemize}
For the complete parts, further features needed were:
\begin{itemize}
\item Protection of the parts from lab air, or during transport, by storage in purged double-bags or containers.
\item For the instrument, continuous gas-purge of the unit with clean gas (dry N$_2$), to prevent ingress of contaminated lab air (this purge is maintained until launch).
\item A separate continuous purge of the detector assembly with dry high-purity nitrogen gas, to protect the KBr photocathode coatings from exposure to humidity.
\item Precautions in vacuum-tests, to prevent out-gassed products from re-deposition on optics, especially during the process of warm-up and venting of the chamber at the end of a test.
\item Verification of the cleanliness levels throughout the build and test phase, by monitoring the environment using witness samples. For molecular contaminants, these were optical plates that were measured using the ESA  FTIR spectrum method (with sensitivity to key contaminant species of \textasciitilde 50\,ng/cm$^2$). For particulates, particle-fall-out plates were used. These were used for regular monitoring of the clean room, and for each specific test. In addition, the vacuum test facilities and all bake-outs had to be verified using quartz crystal microbalance (QCM) measurements.   
\end{itemize}
These cleanliness precautions are at similar stringency to those in ultra-high-vacuum experiments in other science areas using UV (e.g.\ synchrotron facilities).

\subsection{Subsystems testing}
For several of the subsystems, performance and calibration tests were made by the supplier, at unit-level. This was in part to allow the tests to be made early to reduce risk, and also in some cases because more detailed characterisation could be done at this level. The main examples are:
\begin{itemize}
\item Mechanisms: for scan-focus, and slit change, detailed metrology of the mechanism responses, such as stability, repeatability and thermo-mechanical effect
\item Optics (mirror, grating and photo-cathode) spectral responses
\item Detector assembly, functional and UV tests, and environment test, especially thermal as detector cooling is a critical requirement
\end{itemize}

\subsection{Instrument-level testing}
\subsubsection{Overview of test scheme}
\begin{figure*}
\includegraphics{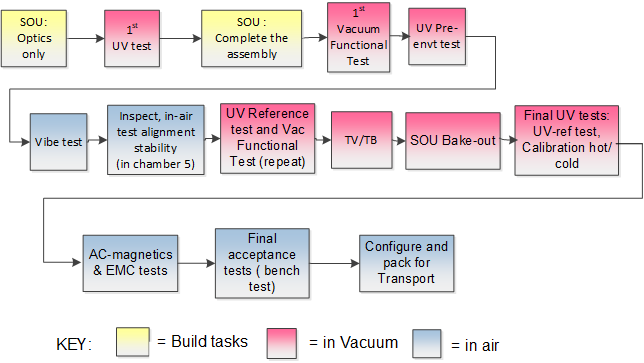}
\caption{Flow diagram of instrument-level tests.}
\label{fig:inst_test_flow}
\end{figure*}

The different stages of instrument-level testing are given in the flow diagram in Fig.~\ref{fig:inst_test_flow}. The main features are:
\begin{enumerate}
\item The first instrument-level test was the `first VUV test' for which the instrument SOU was only part-built. This test was needed immediately after the optics build (and is described in Sect.~\ref{sect-perf}), and before completion of the rest of the SOU parts. This is because it was anticipated that alignment-adjustment might be needed after these first VUV images were obtained. This test is considered as instrument-level as it uses the same system set-up as the subsequent VUV tests.
\item The SOU environmental tests (vibration and thermal-vacuum), are made on the final build of the unit. A VUV reference test, meaning a set of tests to verify all the sensor performances, with the instrument in the vacuum-test chamber, was made before the vibration test, in between vibration test and thermal-vacuum and thermal-balance (TV/TB) test, and again after TV/TB.
\item The instrument has extensive bake-out (out-gassing) test requirements, and the set-up required for this is similar to that for TV-test. Likewise the final calibration tests require the full temperature range of the instrument. For this reason, and to save time as well as minimise handling, all four tests (post-vibe reference test, TV/TB, bake-out, and calibration) were planned into the same vacuum-test procedure. The approximate durations of each phase were: TV/TB: 2 weeks, bake-out: 1 week, calibration: 1 week.
\item After completion of final calibration, the SOU was removed from the test-chamber, and the remaining tests done were bench tests, for functional, EMC and magnetics testing.
\end{enumerate}
With this scheme, there were three test-campaigns in the vacuum chamber facility in total (see Fig.~\ref{fig:inst_test_flow}).

\subsubsection{Vacuum test facility set-up}
\label{sect-env-vacuum}
\begin{figure}
\includegraphics[width=\linewidth]{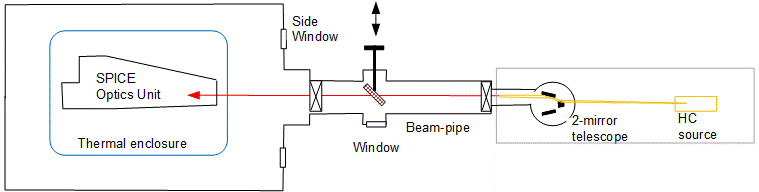}
\caption{Instrument calibration: Test set-up with hollow-cathode (HC) source at the Physikalisch-Technische Bundesanstalt, Berlin.}
\label{fig:vacuum_test}
\end{figure}

A schematic of the facility is shown in Fig.~\ref{fig:vacuum_test}. The hollow-cathode (HC) based calibration source was originally developed for use on the SOHO/CDS instrument \citep{soho:lab_cal}, in collaboration with the Physikalisch-Technische Bundesanstalt (PTB, Berlin), and later used for the Hinode/EIS instrument, where a detailed description of its use was given \citep{solarb:lab_cal}.

The solar instrument test chamber is a 1-metre diameter cylinder tank, in which the SPICE optics unit is positioned as shown, with the optical axis along the chamber axis. The flight harnesses are included on the unit, but for reasons of molecular cleanliness (out-gassing) the Electronics Box is positioned outside the chamber, using harness extensions through chamber feedthroughs. 

The VUV test source used has been described previously by \cite{soho:lab_cal}. It is a collimated test beam produced by the hollow-cathode plasma discharge lamp \citep{soho:radiometric_cal}, which has an exit aperture of 0.6\,mm diameter placed at the focal-point of a 2-mirror grazing-incidence telescope acting as a collimator.
For use in the test, the source vacuum-system has to be coupled to the instrument chamber via a beam-pipe; this is temporarily installed for these tests and so is isolated from the instrument and source chambers via gate-valves. In this configuration it is not possible to view directly either the instrument aperture, nor its alignment-cube, and this is an obstacle to setting up the source, and to determining the instrument pointing (detector line-of-sight relative to the cube), which is an important part of the test aims. To overcome this and align the system, the chamber uses a viewport which is off-axis (adjacent to the beam pipe), to allow an optical alignment reference attached to the instrument bench to be viewed. The alignment of this reference relative to SPICE then has to be measured before the chamber is closed. In addition, the beam-pipe includes a deployable 45$^\circ$ mirror, which can be inserted in order to obtain a visible-light view into the instrument (while blocking the VUV path, however).

\subsubsection{Vibration test}
The vibration test campaigns on the SPICE SOU Flight Spare (FS) and Flight Model (FM) were performed and successfully completed in March 2016 and February 2017, respectively. During the FS vibration test an additional optical alignment measurement was carried out on the non-adhesive mirror mount in order to verify the mechanical stability for each load case. During both test campaigns the instrument was double bagged and purged to comply with cleanliness and contamination control requirements.  
Responses from the applied accelerometers (19 for FS, 13 for FM) were acquired, stored and analysed for each load case to verify the applicable requirements in terms of structural integrity and minimum resonance frequency as well as to improve correlation with the numerical models (both NASTRAN and ANSYS finite elements models). The tests were carried out in the order foreseen by the test procedure, namely:
\begin{itemize}
\item Initial sine survey
\item -12\,dB bedding-in random vibration
\item Bedding-in sine survey
\item Random vibration
\begin{itemize}
\item -12\,dB, -9\,dB, -6\,dB, -3\,dB, sine survey, 0\,dB
\end{itemize}
\item Intermediate sine survey
\item High-level sine
\item Final sine survey.
\end{itemize}
During the FM test campaign, for the Z axis testing only, an addition -6\,dB random vibration run was conducted in order to verify the effectiveness of the notching applied to protect the SCM unit. 

\subsubsection{Thermal-vacuum and thermal-balance test}
To implement the needed thermal cases, the instrument was positioned on a mounting plate, and within a thermal enclosure, both of which are heated/cooled by a fluid loop heat exchanger. The cooling for the detector cold-element is by a separate fluid loop.    

The thermal test campaign consisted of a thermal vacuum test and a thermal balance test. The objectives of these tests were to demonstrate the operation of the SOU across the predicted flight temperature range, to provide environmental stress screening and to provide well-characterised data for validation and correlation of the thermal model. 

All the objectives of the test were ultimately satisfied. In addition, the test provided useful information regarding many operational aspects of the instrument, such as the PID control of the heaters used within the detector assembly. The data acquired during the thermal balance test enabled the thermal model to be correlated to a high level of accuracy, thus giving confidence to the flight thermal predictions that were subsequently produced.

\subsubsection{Vacuum functional tests}
The functional elements of SPICE (motors, detectors, high voltage supplies, etc.) are only fully operational in a vacuum environment.  In order to demonstrate this functionality before, during and after environmental testing (vibration and thermal vacuum tests), full functional tests were carried-out to exercise all capabilities of the instrument under vacuum.  Reduced tests were carried-out in the hot and cold phases of the thermal vacuum test, to demonstrate compliance with requirements on the maximum and minimum operating temperatures of the instrument.  The tests were also designed to provide engineering calibration data, such as the sensor reading versus angle relationship for the scan mirror.  

All tests were successful, demonstrating that SPICE functionality was unaffected by exposure to the instrument level vibration and in-flight thermal environments.  The tests also allowed the team to develop and debug the operating sequences required for in-flight commissioning, such as opening the doors and switching on the detectors for the first time.  

\subsubsection{Magnetics tests}
The main requirements for these tests were to show sufficiently low field effects (on the other Solar Orbiter instruments) for:
(a) Magnetic dipole (i.e.\ direct-current effect), due in particular to the motors in the SPICE mechanisms, which have a permanent dipole, even when not operating; and (b)
AC magnetic fields, due to motor operations and devices in the electronics box.

For the magnetic-dipole, the main measurements were at subsystem level on FM or FS parts, and made at the National Physical Laboratory (NPL, London, UK). For the AC-magnetics, the measurements were at instrument-level on the optics unit (EM model) and electronics box (FS model), made at the same test house as used for EMC testing. These measurements were with AC magnetometers placed at different locations and distances around the SPICE units.

The dipole results were within specifications for all subsystems apart from the Slit Change Mechanism (SCM). This result matches the EM model, where it was identified that magnetic shielding was required for the SCM.  The shielding did not perform as well as anticipated, and the overall dipole results for the instrument were subject to a waiver. However, since the dipole is constant, this can be accounted for in the operations of the magnetometer instrument.  

The AC results showed that the SCM and focus motors generate emissions at 25\,Hz, which was expected.  Future spacecraft testing will determine the level of interference that this causes, but it is not expected to be a significant issue. The timing of any motor operation is recorded by the SPICE flight software, allowing the interested parties to identify the source of the disturbance, and to account for it in their own measurements. This will minimise the impact on science operations.  

\subsubsection{EMC tests}
The main EMC tests were performed on the EM SEB and optics unit.  A full suite of susceptibility and emissions tests was carried-out at a test house (Element, Dorset, UK), which provided the anechoic chamber and equipment needed to meet the required test standards.  Further tests were done on the FM and FS hardware, in agreement with ESA. A subset of conducted emissions and other electrical tests was performed on the FM optics unit at RAL, since the test house could not meet the required cleanliness standards.

The susceptibility testing was very successful, and provided a high level of confidence that SPICE will not be disturbed by emissions from other Solar Orbiter instruments or the spacecraft itself. Emissions testing on the EM identified several areas of concern, and updates to the FM design were implemented based on the lessons learned. The FM results confirmed that motor operations will be considered as `EMC noisy', but normal science studies (with mirror scanning) are not expected to have any significant impact on the rest of the spacecraft. As with the magnetic testing, future tests at spacecraft level and in-flight will characterise this further.  

\section{Characterisation and calibration of performance}
\label{sect-perf}
\subsection{Subsystem testing}
\subsubsection{Detector subsystem VUV tests}
Because the detector assembly is a new development, and its performances are central to the final performance of the instrument, extensive tests were made at detector level. These were aimed at verifying these aspects:
\begin{itemize}
\item Spatial resolution: by recording images of test-targets and various size pinholes on each detector, including through-focus.
\item Radiometric calibration: by including in the set-up a calibrated VUV photo-diode, to measure the beam power, and combining with detector quantum efficiency measurements made on the micro-channel plate wafer prior to detector assembly (Table \ref{table:mcp_efficiency}).
\item Detector throughput: performing radiometric calibration measurements at a variety of MCP voltages allows the optimal voltage to be determined.
\item Linearity: controlling the intensity of light incident on the detector, combined with varying the exposure time, determines the response curve.
\item Flat-field: the detector is mounted on an $x$-$y$ stage so that images can be taken versus varying lateral position, in order to allow the effect of beam-shape to be removed from the image data to produce the true flat-field response.
\item Noise: the statistical properties of the detector are explored by taking multiple images under the same observing conditions.
\item Dark signal: Images are taken at varying exposure times with no light, to assess the thermally generated signal in the APS as a function of temperature.
\end{itemize}
The tests were made on different models of the detector; first on early prototypes, then on EM and finally on FM.

\begin{table}
\caption{Detector quantum efficiency (QE) measured on the micro-channel plate wafer prior to detector assembly. Results are given for short-wavelengths (SW) and long-wavelengths (LW) channel.}
\centering
\begin{tabular}{SSS}
\hline\hline
\multicolumn{1}{c}{{$\lambda$}} & \multicolumn{2}{c}{{QE (\%)}} \\ 
\multicolumn{1}{c}{{(nm)}} & \multicolumn{1}{c}{{SW}} & \multicolumn{1}{c}{{LW}}\\
\hline 49.0 & 22.8 & 22.2\\
       58.0 & 23.6 & 23.4\\
       73.7 & 8.7 & 8.8\\
       83.4 & 17.2 & 17.5\\
       93.2 & 21.1 & 18.7\\
       99.0 & 22.8 & 20.6\\
       104.8 & 24.9 & 21.9\\
       120.2 & 19.5 & 17.6\\
       174.7 & 0.22 & 0.28\\
\hline
\end{tabular}
\label{table:mcp_efficiency}
\end{table}

In the detector test set-up, developed and made by GSFC, the Detector Assembly is placed in the RAL test chamber, and illuminated with a test beam from a commercial krypton lamp (123\,nm). The lamp output is collimated to a beam diameter of approximately 1\,cm, and this illuminates the detector either directly (to fill the array for flat-field tests), or via a focusing mirror (for imaging tests). The collimator has a series of exchangeable field-stop targets (pinhole apertures) to image on to the detector, and also a series of pupil apertures to vary the intensity (imaging f-number). These detector-level tests cover the properties discussed above.

\begin{figure}
\includegraphics[width=\linewidth]{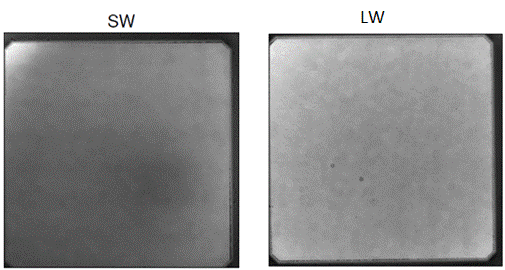}
\caption{Flat-field (relative responsivity across the arrays, shown in grey-scale), for SW and LW detectors, in the APS image (1024$\times$1024\,pixels), measured at 123\,nm. The dark borders occur because the detector fibre-optic coupler does not fully fill the APS.}
\label{fig:flat_field}
\end{figure}

For the FM, the final flat-field patterns are as shown in Fig.~\ref{fig:flat_field}.  Each detector shows a hexagonal fine scale structure due to the micro-channel plate intensifiers and the fibre optic blocks, with a 1\,$\sigma$ variance of 1.6\% for the SW channel, and 2.3\% for the LW channel.  In addition, there are large scale differences in the response which range from about 0.8 to 1.5 for the SW channel, and 0.9 to 1.3 for the LW channel, relative to the median, with the highest values near the detector edges.  Both the large and small scale variations in the flat field are removed during standard data processing.

\subsection{Instrument-level optical tests}
\subsubsection{Telescope optical tests}
The single mirror telescope consists of the entrance-aperture, the mirror and the SPICE slit. The telescope optical performance and alignment to instrument cube was tested in visible light during the instrument build. For the image-quality test, a wavefront error (WFE) measuring interferometer was used, probing the telescope in double-pass, by using a reflective-sphere located at the slit position, for the on-axis FOV position. 

\begin{figure}
\includegraphics[width=\linewidth]{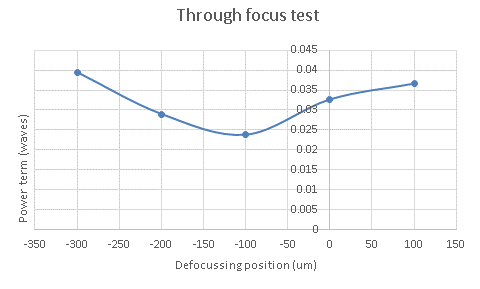}
\caption{Telescope through-focus wavefront error (focus term), showing that optimum focus is at \textasciitilde 100\,\SI{}{\micro\metre} from mid-range (range of mechanism is $\pm$\,500\,\SI{}{\micro\metre}).}
\label{fig:through_focus}
\end{figure}

In final alignment, the  telescope achieved a WFE of \textasciitilde 0.2 waves peak-to-valley at 633\,nm (centre of FOV, best-focus), which is adequate for required spatial image quality. 
The through-focus WFE was also measured, to verify that the telescope is set at best-focus (at mid-range of focus-mechanism), and the result is shown in Fig. \ref{fig:through_focus}.

\subsection{VUV tests}
\label{sect:vacuum_tests}
\begin{figure}
\includegraphics[width=\linewidth]{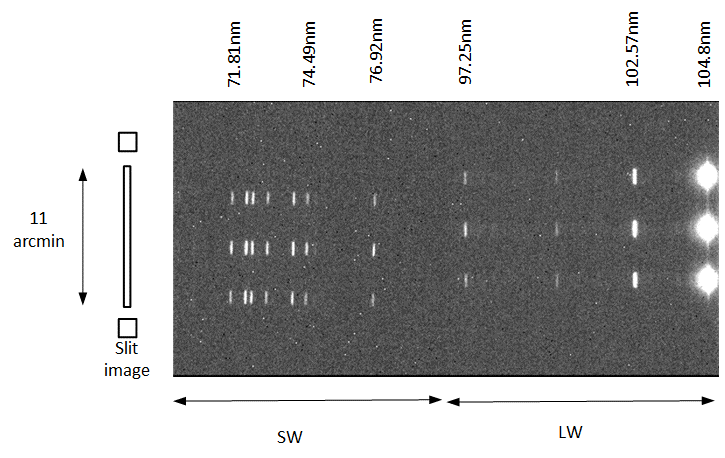}
\caption{Composite SPICE spectral image (argon spectrum), of the three beam positions used. On the left, the slit image is displayed on the same scale. The identification of the hollow-cathode-source argon lines wavelengths is derived from the literature. It should be noted that the slit images appear at different row positions on the SW and LW detectors. }
\label{fig:vacuum_uv}
\end{figure}

The tests for VUV performance were performed in the final test setup in a thermal-vacuum chamber, as described in Sect.~\ref{sect-env-vacuum}. For this the test-beam was a collimated VUV calibration source originally developed for use on the SOHO/CDS instrument \citep{soho:lab_cal}. This source has some geometric limitations for SPICE testing, in that it has (a) a small aperture of  \textasciitilde 5\,mm diameter, compared to the SPICE aperture of \textasciitilde 43\,mm, and (b) an angular range of \textasciitilde 1.7\arcmin diameter, which is larger than the SPICE angular resolution of \textasciitilde 4\arcsec.
The source was operated to obtain the VUV emission line spectrum of argon, known lines of which occur in the SPICE spectral range. The instrument performance was characterised at three beam positions in the along-slit FOV, as shown in Fig.~\ref{fig:vacuum_uv}. 

\begin{figure}
\includegraphics[width=0.75\linewidth]{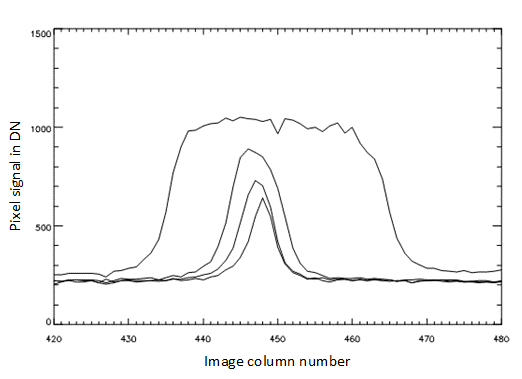}
\includegraphics[width=0.24\linewidth]{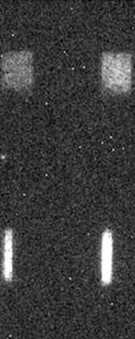}
\caption{Measured spectral-line profiles of all 4 slits (2\arcsec, 4\arcsec, 6\arcsec, and 30\arcsec), at 74.03\,nm wavelength (detector column \textasciitilde 450).}
\label{fig:spectral_profile}
\end{figure}

To verify that SPICE meets its spectral resolution requirement, the 2\arcsec\ line images were analysed, for their LSF width (FWHM) versus spectral position (different argon lines), FOV position, and instrument temperature. It is found that the instrument meets its requirement of 4-pixel FWHM over most of these ranges, and the worst-case value at the edge of the FOV is \textasciitilde 4.5\,pixels. The LSF was also checked versus slit number, and the plot in Fig.~\ref{fig:spectral_profile} shows the spectral line profile of all 4 slits overlaid.

Other tests were
%
imaging tests versus temperature (instrument hot and cold operational cases), and a test of radiometric sensitivity, made by scanning the wide slit across the test beam, to measure total beam signal. This was combined with the known source beam power (in the argon lines 71.8 to 74.5\,nm, for which PTB have a power measurement).

\subsection{Wavelength Calibration}
The wavelength calibration of SPICE has been derived using data described in Sect.~\ref{sect:vacuum_tests}.

The measurements have been taken at room temperature (with a grating temperature of $18.9$\textdegree C) for the four slits. Exposures of 300\,s and 120\,s were required to produce suitable spectra for the three narrower slits and the 30\arcsec\ slit respectively. Spectral images were formed around the middle of the slits and spectral intensity curves were made by averaging over 29\,pixels (SW) and 39\,pixels (LW) along the slit length. The spectral line profiles were fitted using Gaussian shapes to derive the centroids and the FWHM. 

The pixel-to-wavelength calibration was made for the 2\arcsec\ slit by comparing the list of measured lines and their centroids against a list of standard lines \citep{spectra_database}, using a linear fit. The dispersion derived by the fit coefficients was 0.009562\,nm/pixel for the SW and 0.008307\,nm/pixel for the LW channels. These values were compared to the design values of 0.009515\,nm/pixel and 0.0083\,nm/pixel, showing satisfactory agreement.

The wavelength ranges established by this wavelength calibration are shown in Table \ref{table:wavelength_ranges} and are well within the requirements. Only the effective useable pixels are included to determine the wavelength ranges in Table \ref{table:wavelength_ranges}. These are 10 to 975 for SW and 35 to 1003 for the LW detectors, as derived from flat field measurements.

\begin{table}
\caption{Measured SPICE wavelength ranges and related dispersion.}
\begin{tabulary}{\columnwidth}{LLLL}
\hline\hline Detector & Nominal Range (nm) & Measured / Useable Range (nm) & Derived Dispersion (nm/pixel)\\
\hline SW~~~~~~~~~~~~ & 69.8593 – 79.5931 & 69.7008 – 78.9280 & 0.009562\\
\hline LW~~~~~~~~~~~~ & 96.8704 – 105.361 & 96.8783 – 104.919 & 0.008307\\
\hline
\end{tabulary}
\label{table:wavelength_ranges}
\end{table}

This calibration is also valid for the 4\arcsec\ and 6\arcsec\ slits. The 30\arcsec\  slit is much wider and contains also spatial information, but the results are still consistent with the other slits.

The wavelength calibration varies with north-south location on the detector due to relative roll of the slit and the detector. The roll angles are 0.017\,radian for the SW, giving a maximum shift of 5\,pixels in the spectral direction at each end of the slit, and 0.007\,radian for the LW giving a slightly smaller shift of 2\,pixels.

VUV measurements were also performed at different grating temperatures to analyse the thermal cycle variations and spectral shift, which will affect the instrument along the orbit. Two main cases are reported: a cold case where the temperature of the grating was  $-33$\textdegree C, and a warm case where the grating temperature reached 51.8\textdegree C. The 4\arcsec\ slit was used for this investigation. 

We have found a shift of around 10\,pixels towards longer wavelengths for the cold case and a shift of around 15\,pixels toward shorter wavelengths for the warm case, which should simulate the Solar Orbiter perihelion approach.  This analysis will be done again in-flight, as the conditions might change and a curve for this shift will be provided along the orbit.

\subsection{Photometric Sensitivity}
The instrument responsivity is quantified in terms of the total detected signal, per incident radiance $L$ (in W/m$^{2}$~steradian) within a chosen instrument spatial-FOV of solid-angle $\Omega_{S}$, where this is given by the product of the angular sizes of the chosen slit width and along-slit binning. This instrument signal in total DN, summed over all pixels, for exposure time $t_{exp}$, is given by:
\[
N = \frac{L}{hv}A_{ape}\Omega_{S}R({\lambda})t_{exp} \, .
\] 
Here $A_{ape}$ is entrance aperture area. $R(\lambda)$ is the instrument responsivity, in detected signal DN per photon entering the instrument etendue $A\Omega$.

Theoretically $R(\lambda)$ is given by:
\[
R(\lambda) = R_{mir}(\lambda)\,\eta_{gra}(\lambda)\,QDE_{det}(\lambda)\,g_{det} \, ,
\]
where $R_{mir}$ is the mirror reflectivity, $\eta_{gra}$ is the grating absolute efficiency, $QDE_{det}$ is the detector photocathode quantum detection efficiency (electrons created per incident photon) and $g_{det}$ is the detection system gain (DNs per photo-cathode electron). These efficiency parameters of each component or subsystem are separately measured during its development, for performance verification.

The on-ground radiometric calibration at instrument-level then has the aim of measuring the overall responsivity $R(\lambda)$. In ground calibration of the FM, this is made using the hollow-cathode source that generates several spectral lines of argon, for which the total power $P$ in the test beam is approximately known (from characterisation of the source on previous projects). In the above equation, for this test the known power $P$ replaces the product $LA_{ape}\Omega$. In obtaining the total signal $N$ in this test, it is necessary to integrate not only over the whole SPICE image plane, but also over multiple images, while scanning the SPICE pointing across the test beam. This is because the angular size of the test beam is approximately 2\arcmin, which is larger than what can be captured with the widest SPICE slit (0.5\arcmin).

This knowledge of source power is so far only for the argon lines in the range $\lambda$ = 71.8 to 74.5\,nm. In order to improve the accuracy of this calibration, and to extend it further across the SPICE range, the test source was returned to the Metrology Light Source (MLS) located near the BESSY II synchrotron for re-calibration.
\begin{figure}
\includegraphics[width=\linewidth]{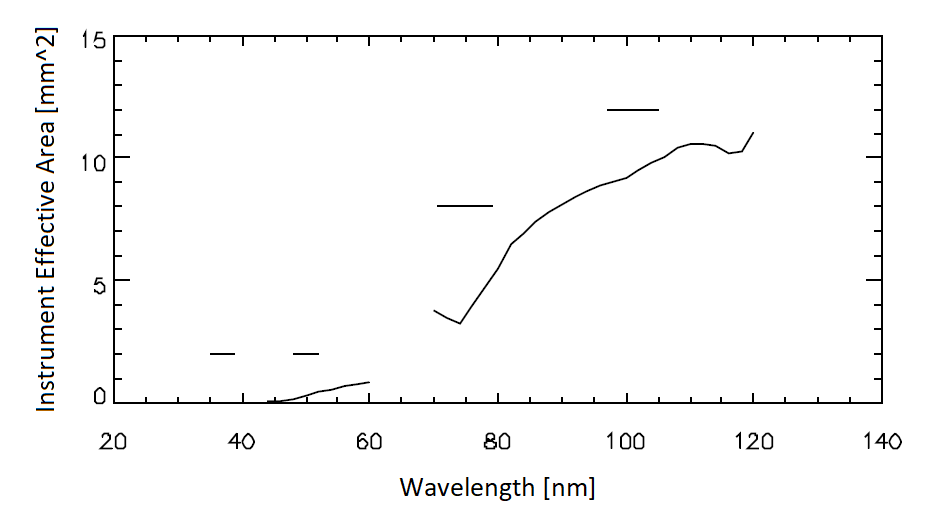}
\caption{SPICE effective area, derived from the measured performance of each subsystem.  The horizontal bars show the wavelength coverage of the SW and LW channels for second order (lower left) and first order.}
\label{fig:effective_area}
\end{figure}

The expected performance in terms of effective area (as derived from subsystem level measurements) is shown in Fig. \ref{fig:effective_area}.  The first order responsivity is mainly a function of $QDE_{det}$, while the second order responsivity is strongly affected by the grating efficiency.  There is no responsivity in the SW channel for second order.  

\subsection{Spectral Hybrid Compression}

The Spectral Hybrid Compression (SHC) scheme adopted by SPICE is a lossy scheme that takes
full advantage of the similarities of spectral image features in all three independent
dimensions of the data set \citep{DeForest2015}.  SHC works by Fourier transformation of
the spectral axis in a narrow spectral window around
the line of interest, which allows the line profile to be coded as a set of Fourier coefficients.
This coding scheme concentrates the information content of the instrument's spectral images,
into the first few Fourier coefficients of the spectral profile.  In particular, the key
features of the line for plasma diagnosis -- intensity, width, and central wavelength -- are
captured in the zero-frequency amplitude, the amplitude ratios of the first few coefficients,
and the phase relationship of the first few coefficients, respectively.  Spatial redundancy is
exploited also, as the Fourier coefficients themselves are treated as individual image planes
that are subjected to lossy compression via a wavelet transform (similar to JPEG2000).  The
data volume allocated to higher spectral frequency Fourier planes is less than that allocated
to the low spectral frequency planes, accounting for the lower information content of the
higher spectral frequencies.  A metadata tag indicates compression quality of each line
profile; this tag is itself losslessly encoded to minimise impact on the data stream while
still providing a reliable quality indicator to highlight any regions where compression
artifacts may be significant.

Because it is a lossy scheme, SHC was extensively validated in the early design phases of
SPICE development.  An important component of that process was testing via compression 
of several high quality (long exposure) spectral scans from the \textit{Hinode}/EIS
spectrograph, degraded by the addition of Poisson
noise to correspond to the noise characteristics of typical SPICE observations.  The spectral
data sets included both simple line structure and multi-component line structure elements.  
In each case, we compared post-facto line fits of the ``original'' degraded EIS data to
similar fits taken from the compressed-and-expanded data.  A benchmark of compression to 16
bits per profile ($28\times$) was used for validation, compared to the less aggressive 22-28
bits per profile anticipated for flight campaigns.  Typical performance was: (1)
line intensity was reproduced to within a small fraction (typically under 25\%) of the photon
counting noise in typical observations; (2) line centre wavelength (Doppler shift) was
reproduced to within 0.1 pixel RMS across all cases where the line fit converged; (3) line
full-width was reproduced to within 0.2 pixel RMS across all cases where the line fit
converged; (4) sidelobes with intensity ratios as little as 5\% compared to the line core 
were reproduced.  Throughout testing, errors induced by lossy compression proved to be (a)
uncorrelated to the signal and (b) well within the error budget driven by the SPICE science
requirements. 
\section{Operations concept}
\label{sect-ops}

Like for the other instruments of Solar Orbiter, the SPICE scientific observations are planned as elements of a succession of SOOPs (Solar Orbiter Observing Programs) forming the long term plan agreed upon on a per-orbit basis by the Science Working Team (SWT). Each SOOP is designed to address one or more of the scientific objectives described in the Solar Orbiter Science Activity Plan \citep[][]{Zouganelis2019a}. The actions to be performed by the instrument are planned as a timeline, meaning a sequence of telecommands, encapsulated in several BOPs (Basic Observing Programs). A BOP can conveniently group together commonly used sequences of telecommands, like the selection of a slit followed by the execution of a study, or the preparation of the instrument from standby to operating mode. Most of the scientific observations will consist of `studies', as introduced in Sect.~\ref{sec:studies}. The commanding software (or planning tools) implements these three concepts (timeline, BOPs and studies) in three modules,
%
The study generator, the BOP generator, and the timeline tool.

The planning tools are written using the Django Python web framework for the back-end, while the front-end was built with Bootstrap and JQuery. We use Selenium (functional testing) to ensure everything works as expected and pytest (unit testing) to make sure the code doesn't break. We also have a Jenkins instance that fires up at every git commit to launch tests and generates a SonarQube report that gives us insight on what could be improved or refactored. The whole project can be deployed from a Debian Docker image.
The planning tools are accessed via user accounts with hierarchical privileges (observer, planner, administrator) and allows several users to work simultaneously while avoiding conflicts. Observers can design studies, BOPs and timelines. Planners have the additional rights to approve them for transfer to Solar Orbiter's Mission Operations Centre (MOC) or Science Operations Centre (SOC). Administrators have further rights to manage the database. In the following, we give an overview of the main functionalities of the planning tools.

\subsection{Study generator}
\label{sec:study-gen}

\begin{figure*}
\includegraphics[width=\linewidth]{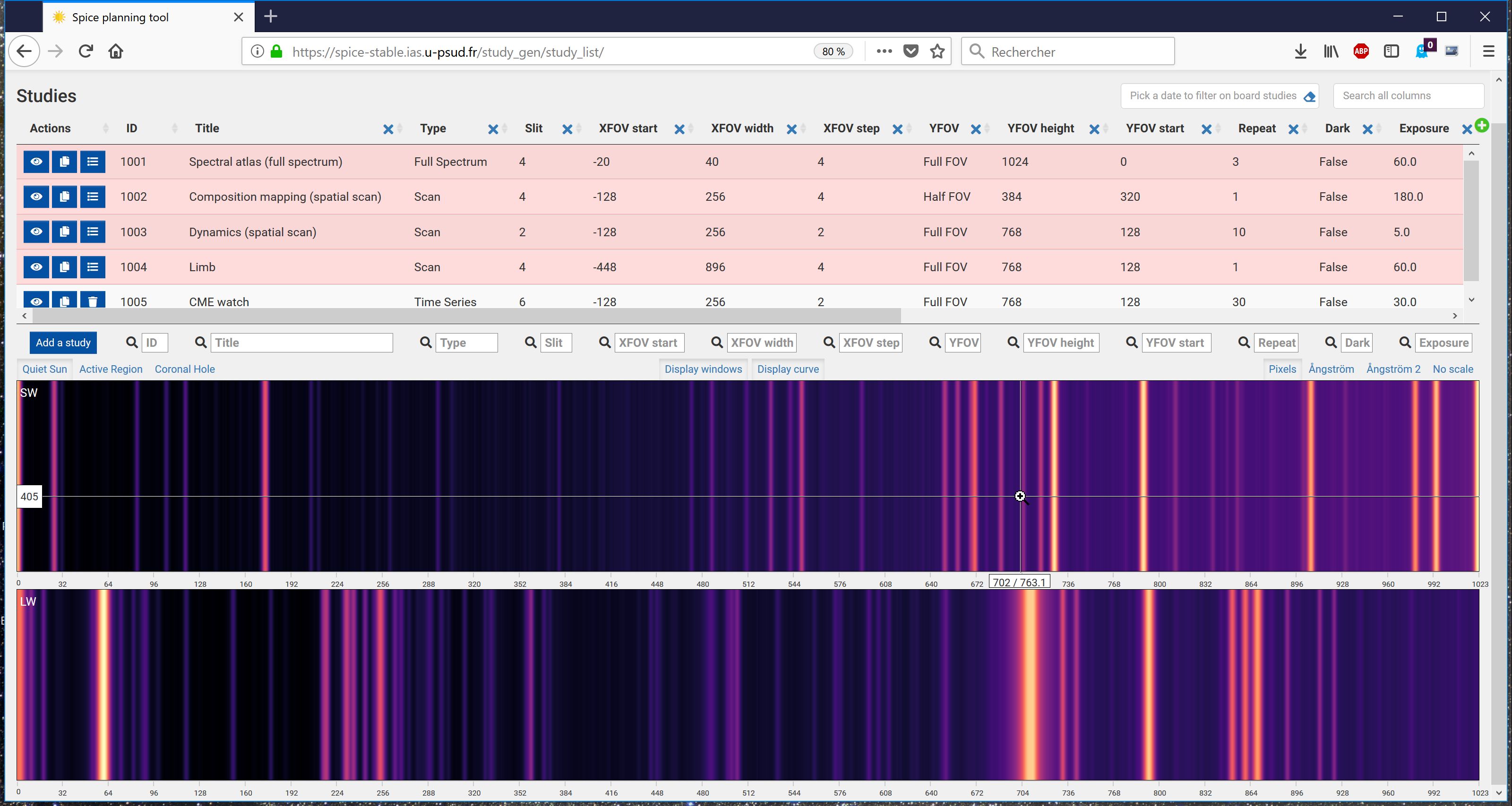}
\caption{The graphical user interface of the SPICE study generator.}
\label{fig:study_generator}
\end{figure*}

The study generator allows a user to design studies compliant with all the instrumental constraints. The top-level element of its graphical user interface (GUI) is the study chooser (Fig.~\ref{fig:study_generator}).
The table lists all the studies present in the database (not only those on board) and their main parameters and allows to search among existing studies.
The two images at the bottom are typical spectra corresponding to the short and long wavelength detectors respectively.
From there the planner can either edit an existing study or create a new one in the study editor.
Studies can be of four types: full spectrum, spatial scan (raster), time series (sit and stare), or scanned time series.
The planner can then set all the study parameters, like exposure times, start, stop and size of the spatial steps, or the spectral windows necessary for the wanted observations. If the desired spectral window does not exist, the planner can create a new one using the window editor.

With the study editor and window editor, the planner is able to set all the parameters of the on-board LUTs (see Sect.~\ref{sec:studies}) necessary to configure SPICE for the desired observations. The study editor and the window editor enforce all known instrumental constraints. For example, exposure times can only take the values predefined in the on-board LUTs; spectral windows can not overlap; for scanned studies, scans must contain a multiple of 32 positions if there are any narrow windows ($<32$\,pixels width) or if SHC compression is used.

The on-board LUTs allow the storing of 64 studies of which 16 are reserved for engineering purposes. These 64 studies all use the same LUTs to define, for example, their spectral windows. Since the LUT that defines the spectral windows has 256 entries, the combined 64 studies must not use more than 256 windows. It is worth noting that several study parameters are defined as attributes of the spectral windows: type (intensity or profile), width, compression, binning. Therefore, observing the same spectral interval with two different binnings requires two different windows and uses two entries of the corresponding LUT. While each study is defined independently, the planning tools must therefore ensure that the 64 studies selected for upload to the instrument fit in the existing number of LUT entries. Conflicts (e.g.\ too many spectral windows) are resolved either {\it a priori} by preventing the planner from making modifications or {\it a posteriori} by asking the planner to make choices. In addition, each study is tested on the software instrument simulator before it can be used in a set of 64 selected for upload.


\begin{table*}
\caption{Examples of SPICE observations. The assumed compression rates are 20:1 for spectral windows (SHC) and 10:1 for intensity windows (marked with a $^*$). In these observations, the raster step in the $x$ direction is equal to the nominal slit width.}
\begin{tabular}{cp{.27\linewidth}cccccc}
	\hline\hline
    Objective & Spectral lines & Slit & FOV & Exp. time & Duration & Data rate & Data volume\\
    \hline
    QS dynamics & 
    \ion{C}{III} 997.0, \ion{Ne}{VIII} 770.3, \ion{H}{I} 1025.7, \ion{O}{II} 718.5$^*$, \ion{O}{IV} 787.7$^*$, \ion{Ne}{VI} 997.2$^*$, \ion{Mg}{IX} 749.5$^*$, \ion{Mg}{VIII} 772.3$^*$, \ion{Mg}{VIII} 782.3$^*$, \ion{O}{VI} 1031.9
    & 2" & $256"\times829"$ & 5\,s & 12\,min & 18\,kb\,s$^{-1}$ & 12\,Mb\\
    Composition & 
    \ion{Ar}{VIII} 713.8, \ion{S}{V} 786.5, \ion{Mg}{IX} 706.1, \ion{O}{VI} 1037.6, \ion{Mg}{VIII} 782.3, \ion{Ne}{VIII} 770.4, \ion{O}{VI} 1031.9$^*$
    & 6" & $954"\times829"$ & 60\,s & 2.6\,h & 2.1\,kb\,s$^{-1}$ & 19.2\,Mb\\ \hline
\end{tabular}
\label{table:example_observations}
\end{table*}

\subsection{BOP generator and timeline tool}

As previously described, BOPs are used as a convenient encapsulation for groups of sequences of telecommands that are  occurring frequently, and the BOP generator provides a simple interface for assembling and manipulating sequences of telecommands with their formal parameters and relative time delays.

The timeline tool provides an interface for assembling BOPs into timelines. The tool enforces the mission constraints (rolls, EMC quiet periods, etc.) as described in the E-FECS (Enhanced Flight Events and Communications Skeleton) files provided by the SOC. After validation, the tool automatically converts the timeline into XML files describing the corresponding Instrument Operations Request (IOR) to be sent to the SOC. In Table~\ref{table:example_observations}, we provide two examples of observing sequences designed for the study of Quiet Sun dynamics and elemental composition. The Quiet Sun dynamics sequence fits within the maximum data rate allocated to SPICE and could thus be run continuously. This demonstrates the ability of SPICE to provide high-cadence, high-resolution data despite the orbit-related telemetry constraints of Solar Orbiter.
\section{Data processing and tools}
\label{sect-data}

\subsection{SPICE FITS files} 
A SPICE FITS file will contain a primary Header/Data Unit (HDU), and it may contain one or more additional HDUs (image extensions). All primary and image extension HDUs (observational HDUs) will contain observational data with complete, self-contained headers, meaning there is no distinction between primary HDUs and image extensions other than those required by the FITS standard. In addition, the file may contain a binary table extension that will hold the values of FITS keywords that vary with exposure number. Every observational HDU in LL01, L1 and L2 files (see sections below for details) will store data from a single readout window, dumb-bell region or intensity window, or from a single detector in the case of a full frame readout. The data cubes of all SPICE FITS files will be 4-dimensional, with dimensions {\tt (X,Y,dispersion,time)}. One or more of the dimensions may be singular, for example raster scans have a singular {\tt time} dimension and sit-and-stare observations have a singular {\tt X} dimension. Each observational HDU of an L3 FITS file will contain a single derived data product (e.g.\ intensity, velocity).

\subsection{Low latency telemetry processing}
The SPICE Low Latency Pipeline will run in a virtual machine at SOC, with low latency telemetry data as input, and FITS files of level LL01 as output.  LL01 files will contain uncalibrated data expressed in engineering units. The time will be given in on-board time and the pointing will be relative to the Solar Orbiter boresight. The SOC will provide a simple web-based visualisation tool for the low latency data for observation planning purposes. The LL01 files will be stored in the Solar Orbiter Archive. 

\subsection{Science data telemetry processing}
The SPICE Science Data Pipeline will run in Oslo, with telemetry data as input, and files of multiple levels and file formats as output. All output files from the Science Data Pipeline will be stored in the Solar Orbiter Archive. 

\subsubsection{Level 1: uncalibrated data}
Level 1 FITS files (L1) will be uncalibrated data expressed in engineering units. Production of L1 files will include time conversion from on-board time to UTC and transformation of coordinates from being given relative to the spacecraft boresight to being relative to Sun centre. L1 files will also include additional metadata taken from the study definition database and the timeline. This information will typically be strings that either describe all files observed with a particular study, or that describe a particular instance of a study.

\subsubsection{Level 2: calibrated data}\label{L2}
Level 2 FITS files (L2) will be calibrated files that are ready for scientific analysis: the data will be calibrated to physical units, and will be corrected for flat-field, dark current etc. Geometric corrections that account for slit tilts, spectral slant, detector misalignments, non-uniform dispersion, and other geometric distortions will be applied to the data cubes by interpolation onto a regular grid. The rotation of the field of view due to spacecraft roll will not corrected for, instead it will be described by the {\tt PCi\_j} transformation matrix of the World Coordinate System (WCS) coordinates.

\subsubsection{Level 3: higher level data products}\label{L3}
Level 3 files (L3) will come in different types and file formats:  regular L3 FITS files, concatenated L3 FITS files, and quicklook L3 files.
Regular L3 FITS files will contain data products that are derived from a single L2 file: line intensity, velocity and line width resulting from  line fitting, and other data products derived from the fit parameters, like densities, temperatures, and FIP bias. Each HDU of an L3 file will contain a single derived data product for a single automatically detected and identified emission line. It should be noted that the pipeline may detect more than one emission line in each readout window. 
Concatenated L3 FITS files will contain time series of derived data products, meaning a single concatenated L3 FITS file contains the derived data products stemming from multiple regular L3 FITS files (and hence stemming from multiple L2 files). 

L3 files can also be JPEG images or MPEG movies meant for quicklook purposes. Quicklook L3 files will not contain the metadata that are present in L1, L2, and L3 FITS files. A quicklook L3 JPEG file will be either an image of a derived data product, a wide slit image, or a dumb-bell image. A quicklook L3 MPEG file will be either a movie of a derived data product, or a movie of wide slit/dumb-bell multi-exposure observations. 
 
 \subsection{Manual processing from Level 1 to Level 3}\label{manual}
The processing software, written in IDL, will be provided via SolarSoft. The software will enable users to manually process the L1 data to L2 data by applying calibrations, and to convert L2 to L3 by performing line fitting and creating data products derived from the fit parameters. During manual processing from L1 to L3, the user will be able to choose which of the steps described in Sects.\,\ref{L2} and \ref{L3} should be applied or not, and to tweak the parameters involved in each step. If the user chooses to omit the step performing the geometric correction on the data cube, the geometric distortions will instead be fully described by FITS WCS keywords. 

\subsection{Visualisation and analysis tools}\label{GUI}
Visualisation and analysis tools written in IDL will be provided. The visualisation tools (GUI) will allow for inspection of FITS file data products at all levels. The analysis tools will also provide an interface to the manual processing from L1 to L2 and L3 data, as described in Sect.~\ref{manual}. It is also foreseen to generate browse data for multi-instrument visualisation with JHelioviewer \citep{Muller:2017dq}.

\section{Summary}
The SPICE instrument is a high-resolution imaging spectrometer onboard the Solar Orbiter mission, operating at EUV wavelengths from 70.4$-$79.0\,nm and 97.3\,nm$-$104.9\,nm.
It will provide the quantitative knowledge of the physical state and composition of the plasmas in the solar atmosphere, in particular investigating the source regions of outflows and ejection processes which link the solar surface and corona to the heliosphere. By observing the intensities of selected spectral lines and line profiles, SPICE will derive temperature, density, flow and composition information for coronal plasma in a wide temperature range.

SPICE is of particular importance for establishing the link between remote-sensing and in-situ measurements as it is uniquely capable of remotely characterising the plasma properties of source regions, which can directly be compared with in-situ measurements taken by the Solar Wind Analyser (SWA) instrument suite.
In magnetically closed regions, SPICE will play an essential role in characterising the turbulent state of the plasma over a wide range of temperatures from the chromosphere into the hottest parts of the corona. This is essential to understand which processes heat the plasma and drive the dynamics we observe.

\begin{acknowledgements}
The development of the SPICE instrument has been funded by ESA member states and ESA (contract no.\ SOL.S.ASTR.CON.00070).
The German contribution to SPICE is funded by the Bundesministerium f\"ur Wirtschaft und Technologie through the Deutsches Zentrum f\"ur  Luft- und Raumfahrt e.V. (DLR), grants no.\ 50  OT 1001/1201/1901.  
The Swiss hardware contribution was funded through PRODEX by the Swiss Space Office (SSO).
The UK hardware contribution was funded by the UK Space Agency.
The work of S.K.~Solanki has been partially supported by the BK21 plus program through the National Research Foundation (NRF) funded by the Ministry of Education of Korea. 
Finally, the authors would like to thank the referee, Dr Shimizu, whose comments help to improve the quality of this paper.
\end{acknowledgements}

%
%

\bibliographystyle{aa}
\bibliography{aamnem99,loops,fits_wcs,ral,SO_Book_SPICE_references,SO_Book_cross_references}

\end{document}